\documentstyle[editedvolume,numreferences,epsfig]{crckapb} 

\newcommand{\xbj}{x_{_B}}

\newcommand{\f}[1]{{\bf Fig.~#1}}
\def\be{\begin{equation}}
\def\ee{\end{equation}}
\def\ba{\begin{eqnarray}}
\def\ea{\end{eqnarray}}

\def\ul{\underline}
\def\z0{\rm Z^0}
\def\as{\alpha_{\rm s}}

\newcommand{\oaa}{{\cal O}(\as^2)}
\newcommand{\oaaa}{{\cal O}(\as^3)}
\newcommand{\oaaaa}{{\cal O}(\as^4)}
\newcommand{\epem}{\rm e^+\rm e^-}
\newcommand{\yc}{y_{\rm cut}}
\newcommand{\amz}{\as(M_{\rm Z^0})}

\def\mz{M_{\rm Z^0}}

\def\d2{D_2}
\def\oq{\char'134}

\def\m2{\mu^2}
\def\q{\rm q}

\def\p{\rm p}

\def\q2{Q^2}
\def\asq{\as (\q2 )}

\begin{opening}
\title{
STANDARD MODEL PHYSICS AT LEP 
}

\author{S. BETHKE}
\institute{Max-Planck-Institut f\"ur Physik \\
           (Werner-Heisenberg-Institut)\\
           80805 M\"unchen, Germany}

\end{opening}

\runningtitle{STANDARD MODEL PHYSICS AT LEP}

\begin{document}

\begin{abstract}
Selected topics on precision tests of the Standard Model of the Electroweak
and the Strong Interaction at the LEP $\epem$ collider are presented,
including an update of the world summary of measurements of $\as$,
representing the state of knowledge of summer 1999.
This write-up of lecture notes\footnote{
Lecture given at the International Summer School \oq{\it Particle
Production Spanning MeV and TeV Energies}", Nijmegen (The Netherlands),
August 8-20, 1999.}
consists of a reproduction of slides,
pictures and tables, supplemented by a short descriptive text and a
list of relevant references.

\end{abstract}

\vspace*{-11.0cm}
\begin{flushright}
MPI-PhE/2000-02 \\
January 2000
\end{flushright}
\vspace*{9.0cm}

\section{Introduction}

The physics of elementary particles and forces determined the development of
the early universe and thus, of the structure of our world today (\f{1}).
According to our present knowledge, three families of quarks and leptons, four
fundamental interactions, their respective exchange bosons  and a
yet-to-discover mechanism to generate particle masses are the ingredients
(\f{2}) which are necessary to describe our universe, both at cosmic as well as
at microscopic scales.

Three of the four forces are relevant for particle physics at small distances:
the Strong, the Electromagnetic and the Weak Force.
They are described by quantum field theories, Quantum
Chromodynamics (QCD) for the Strong, Quantum-Electrodynamics (QED) for the
Electromagnetic and the so-called Standard Model of the unified
Electro-Weak Interactions \cite{books}.
The weakest force of the four, gravitation, is the major player only at large
distances where the other three are, in general, not relevant any more: the
Strong and the Weak Force are short-ranged and thus limited to sub-nuclear
distances, the Electromagnetic force only acts between objects whose net
electric charge is different from zero.

Of the objects listed in \f{2}, only the $\tau$-neutrino ($\nu_{\tau}$), the
Graviton and the Higgs-boson are not explicitly detected to-date.
Besides these particular points of ignorance, the overall picture of
elementary particles and forces was completed and tested with remarkable
precision and success during the past few years, and the data from the LEP
electron-positron collider belong to the major important
ingredients in this field. 

This lecture reviews selected aspects of
Standard Model physics at LEP.  
The frame of this write-up is not a standard and text-book-like presentation,
but rather a collection and reproduction of slides, pictures and tables,
similar as presented in the lecture itself.
Since most of the slides are self-explanatory, the collection is only
accompanied by a short, connecting text, plus a selection of references
where the reader can find more detailed information.

\section{LEP: machine, detectors and physics}

A decade of successful operation of the Large Electron Positron collider, LEP
\cite{lep} (\f{3}), provided a whealth of precision data (\f{4}) on the
electroweak and on the strong interactions, through a multitude of $\epem$
annihilation final states (depicted in \f{5}) which are recorded by
four multi-purpose detectors, ALEPH \cite{aleph}, DELPHI \cite{delphi}, L3
\cite{l3} and OPAL \cite{opal}.

In the phase which is called \oq LEP-I",
from 1989 to 1995, the four LEP experiments have collected a total of about 17
million events in which an electron and a positron annihilate into a $\z0$ which
subsequently decays into a fermion-antifermion-pair (see Figs.~4 and~5).
Since 1995, the LEP collider operates at
energies above the $\z0$ resonance, 
$\sqrt{s} \equiv E_{cm} > \mz \cdot c^2$ (\oq LEP-II"), up to
currently more than 200 GeV in the centre of mass system.
The different final states of $\epem$ annihilations can be measured and
identified with large efficiency and confidence, due to the hermetic and
redundant detector technologies realised by all four experiments.

An example of a hadronic 3-jet event, originating from the process $\epem
\rightarrow\z0 \rightarrow q\overline{q} g$ with subsequent fragmentation of
quarks and gluon(s) into hadrons, as recorded by the OPAL detector (\f{6})
\cite{opal}, is
reproduced in \f{7}.

\section{Precision tests of the Electroweak Interaction}

The basic predictions of the Standard Model of Electroweak Interactions, for
fermion-antifermion production of $\epem$ annihilations around the $\z0$
resonance, are summarised in \f{8} to \f{11}, see \cite{books} and recent
experimental reviews \cite{hbu,quast,mnich} for more details.
Cross sections of these processes are energy (\oq s"-) dependent and
contain a term from $\z0$ exchange, another from photon exchange as well as a
\oq $\gamma - \z0$" interference term (\f{8}). 
Measurements of s-dependent cross sections around the $\z0$ resonance provide
model independent results for the mass of the $\z0$, $\mz$, of the $\z0$
total and partial decay widths, $\Gamma_Z$ and $\Gamma_f$, and of the fermion
pole cross sections, $\sigma_f^0$.

Beyond the lowest order \oq Born Approximation", photonic and non-photonic
radiative corrections must be considered (\f{9}); the latter can be absorbed
into \oq running coupling constants" (\f{10}) which, if inserted into the Born
Approximation, make the experimental observables depend on the masses of
the top quark and of the Higgs Boson, $M_t$ and $M_H$.
Measurements 
of the fermion final state cross sections as well as of
other observables like differential cross sections, forward-backward
asymmetries and final state polarisations of leptons (\f{11}) allow to extract
the basic electroweak parameters.

Combined analyses of the data of all 4 LEP experiments by the \oq LEP
Electroweak Working Group" \cite{ew-combined} provide very precise results
(\f{12}): for instance, due to the precise energy calibration of LEP
\cite{e-calib},
$\mz$ is determined to an accuracy of 23 parts-per-million, and the number of
light neutrino generations (and thus, of quark- and lepton-generations in
general) is determined to be compatible with 3 within about 1\% accuracy.
From radiative corrections and a combination of data from 
LEP-I and LEP-II, $M_t$, $M_H$, the
coupling strength of the Strong Interactions, $\as$, the effective weak mixing
angle $sin^2 \theta_{lept}^{eff}$ and the Mass of the W-boson, $M_W$, can be
determined with remarkable accuracy (except for $M_H$ which only enters
logarithmically).
A list of the most recent results \cite{mnich} is given in \f{13}, where also
the deviations of the experimental fits from the theoretical expectations are
given by the number of standard deviations (\oq Pull").

Graphical representations of some of these results are given in \f{14} to
\f{18}.
The significance of counting the number of light neutrino families,  $N_\nu$,
from the measurement of the $\z0$ line shape, based on ALEPH data from the 1990
and 1991 scan period, is displayed in \f{14}.
The gain in precision of electroweak parameters between 1987, before the era of
LEP, and the LEP results of 1999 is demonstrated in \f{15}, for the values of
the leptonic axial and vector couplings, $g_a$ and $g_v$.

The fit result of the Higgs mass, $M_H$, ist given in \f{16}, calculated using
two different input values for the uncertainty of the hadronic part of
the running QED coupling constant, $\Delta \alpha_{had}$
\cite{jegerlehner,davier}, together with the exclusion
limit from direct Higgs production searches, $M_H > 95.2$~GeV (95\% confidence
level) \cite{mnich}.

The measured cross section for $W$ pair production, $\epem \rightarrow W^+W^-
(\gamma )$, is presented in \f{17}, together with the Standard Model prediction
and two \oq toy models" which demonstrate the importance of the $ZWW$ triple
gauge boson vertex and the $\nu_e$ exchange diagram, see \f{5}.
A summary of the available measurements (top) and indirect determinations,
i.e. through radiative corrections
(bottom), of the
$W$ mass is given in \f{18}.
More results and graphs are available from \cite{mnich} and from the home page
of the LEP Electroweak Working Group \cite{ew-combined}.

\section{Jet Physics and Tests of QCD}

A short introduction to the development of hadron physics, from the discovery of
the neutron to the development of QCD and the experimental manifestation
of gluons, is given in \f{19}.
The basic properties of QCD - in comparison with QED - are summarised in \f{20}.
The energy dependence of the strong coupling
strength $\as$, given by the so-called $\beta$-function in terms of the
renormalisation scale $\mu$ and the QCD group structure parameters $C_f$, $N_c
\equiv C_a$ and $N_f$, is described in \f{21}.

In \f{22}, the anatomy of the process $\epem \rightarrow$~hadrons is
illustrated.
Factorisation is assumed to hold when splitting this process into an
electroweak part (annihilation of $\epem$ into a virtual photon or $\z0$ and
subsequent decay into a quark-antiquark pair), the development of a parton
(i.e. quark and gluon) shower described by perturbative QCD, a hadronisation
phase which can be modelled using various different fragmentation or
hadronisation models, and finally a parametrisation of the decays of unstable
hadrons (according to measured decay modes and branching fractions)
\cite{jetset,herwig,models}.

A list of the most prominent QCD topics covered by the LEP experiments is
given in \f{23}.
For a more detailed introduction to QCD and hadronic physics at high energy
particle colliders see e.g.
\cite{qcdbook}; earlier reviews of QCD tests at LEP can be found in
\cite{annrev,hebbeker,scotland}.

One of the most prominent QCD-related measurements at LEP is the
determination of $\as$ from the radiative corrections to the hadronic partial
decay width of the  $\z0$, which is summarised in \f{24}.
The ratio $R_Z = \Gamma_{had} / \Gamma_{lept}$ is a totally inclusive quantity
which is independent of hadronisation effects, and QCD corrections are
available in complete $\oaaa$, i.e. in next-to-next-to-leading order QCD
perturbation theory \cite{r-nnlo,r-hebbeker}.
The determination of $\as$ from $R_Z$, however, crucially depends on the
validity of the predictions of the Electroweak Standard Model.

The basic principles of the physics of hadrons jets, which are interpreted as
the footprints of energetic quarks and gluons, and the definition of
hadron jets are described in \f{25}.
The most commonly used jet algorithms in $\epem$ annihilations are clustering
procedures as first introduced by the JADE collaboration \cite{jadejet}, and
variants of this algorithm \cite{durhamjet,nlo,bkss,cambridgejet} 
as listed in {\bf Tab.~1}.

For these algorithms, relative production rates of $n$-jet events ($n$ = 2, 3,
4, ...) are predicted by QCD perturbation theory, and are
therefore well suited to determine $\as$ and to prove the energy dependence of
$\as$, see \f{26}. 
In particular, the relative rate of 3-jet events, $R_3$, is predicted to be
proportional to $\as$, in leading order perturbation theory.
Corrections in complete next-to-leading order, i.e. in $\oaa$, are available
for these algorithms \cite{nlo,bkss}.

Hadronisation effects, however, may significantly influence the
reconstruction of jets.
This can be seen in \f{27}, where jet production rates are analysed using QCD
model (Jetset) events of $\epem$ annihilation at $\sqrt{s} = 91.2$~GeV before
and after hadronisation, i.e. at parton- and at hadron-level. 
The $purity$ of
3-jet reconstruction, i.e. the number of events which are classified as 3-jet
both on parton- and at hadron-level, normalised by the number of events
classified as 3-jet on hadron level, is displayed in \f{28}.
The energy dependence of hadronisation corrections to measurements of 3-jet
event production rates at fixed jet resolution $\yc$ is analysed in \f{29}.
From these studies, the original JADE and the Durham
schemes emerge as the most \oq reliable" algorithms to test QCD in jet
production from $\epem$ annihilations (for a comparative study of the
newer Cambridge algorithm, see e.g. \cite{camb-studies}). 

Especially the JADE algorithm exhibits small and almost
energy independent hadronisation corrections.
This allows to test the energy dependence of $\as$ and thus, of asymptotic
freedom, without actually having to determine numerical values of $\as$, see
\f{30} \cite{qcd96}.

Hadronic event shapes (\f{31}) are a common tool to study aspects of QCD, and
in particular, to determine $\as$.
For many of these observables, QCD predictions in next-to-leading order
($\oaa$) are available \cite{nlo}, and for some of them, the leading and
next-to-leading logarithms were resummed to all orders \cite{nlla}.

The results of one such study, performed by L3 \cite{l3-as} using
event shapes of LEP-I and LEP-II data plus radiative events at 
reduced centre of
mass energies, is shown in \f{32}, demonstrating the running of $\as$.
For more details on the determination of $\as$ from hadronic event shape and
jet related observables, see eg. \cite{qcdbook,annrev,hebbeker,jade-power}.

A list of high energy particle processes and observables from which significant
determinations of $\as$ are obtained is given in \f{33}.
The most recent measurements, as an update to the world summary of $\as$ from
1998 \cite{radcor}, are listed in \f{34}.

{\bf Table 2} summarises the current status of $\as$ results.
The corresponding values of $\asq$, where $Q$ is the typical hard scattering
energy scale of the process which was analysed, are displayed in \f{35}.
The data, spanning energy scales from below 1~GeV up to several hundreds of
GeV, significantly demonstrate the energy dependence of $\as$, which is in good
agreement with the QCD prediction.

Evolving these values of $\as (Q)$ to a common energy scale, $Q = \mz$,
using the QCD $\beta$-function in $\oaaaa$ with 3-loop matching at the heavy
quark pole masses $M_b = 4.7$~GeV and
$M_c = 1.5$~GeV \cite{4-loop}, results in \f{36}, demonstrating the good
agreement between all measurements.
From the results based on QCD calculations
which are complete to next-to-next-to-leading order (filled symbols in Fig.
36; see also Table 2), a new world average of
$$\amz = 0.119 \pm 0.003\ \hskip1.0cm {\rm [in\ NNLO]}$$
is determined.
The overall error is calculated using a method \cite{schmelling} which
introduces an common correlation factor between the errors of the individual
results such that the overall $\chi^2$ amounts to 1 per degree of freedom.
The size of the resulting overall uncertainty depends on the method
and philosophy used to determine the world average of $\amz$, see
\cite{radcor} for further discussion.

\clearpage


\begin{figure}[h,t]
\begin{center}
\epsfig{file=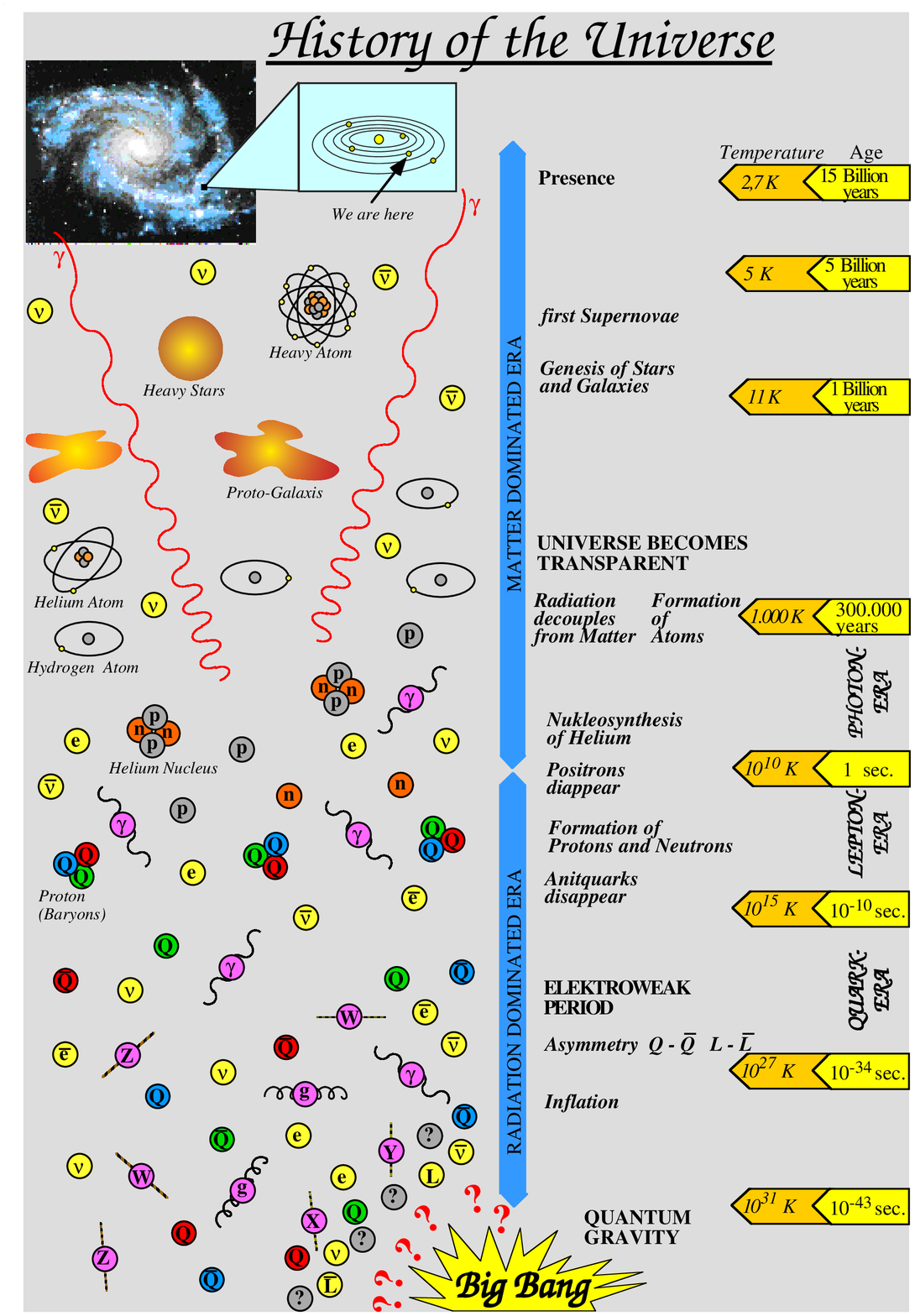,width=12.5cm}
\end{center}
\caption{\label{fig1}
}
\end{figure}

\begin{figure}[h,t]
\begin{center}
\epsfig{file=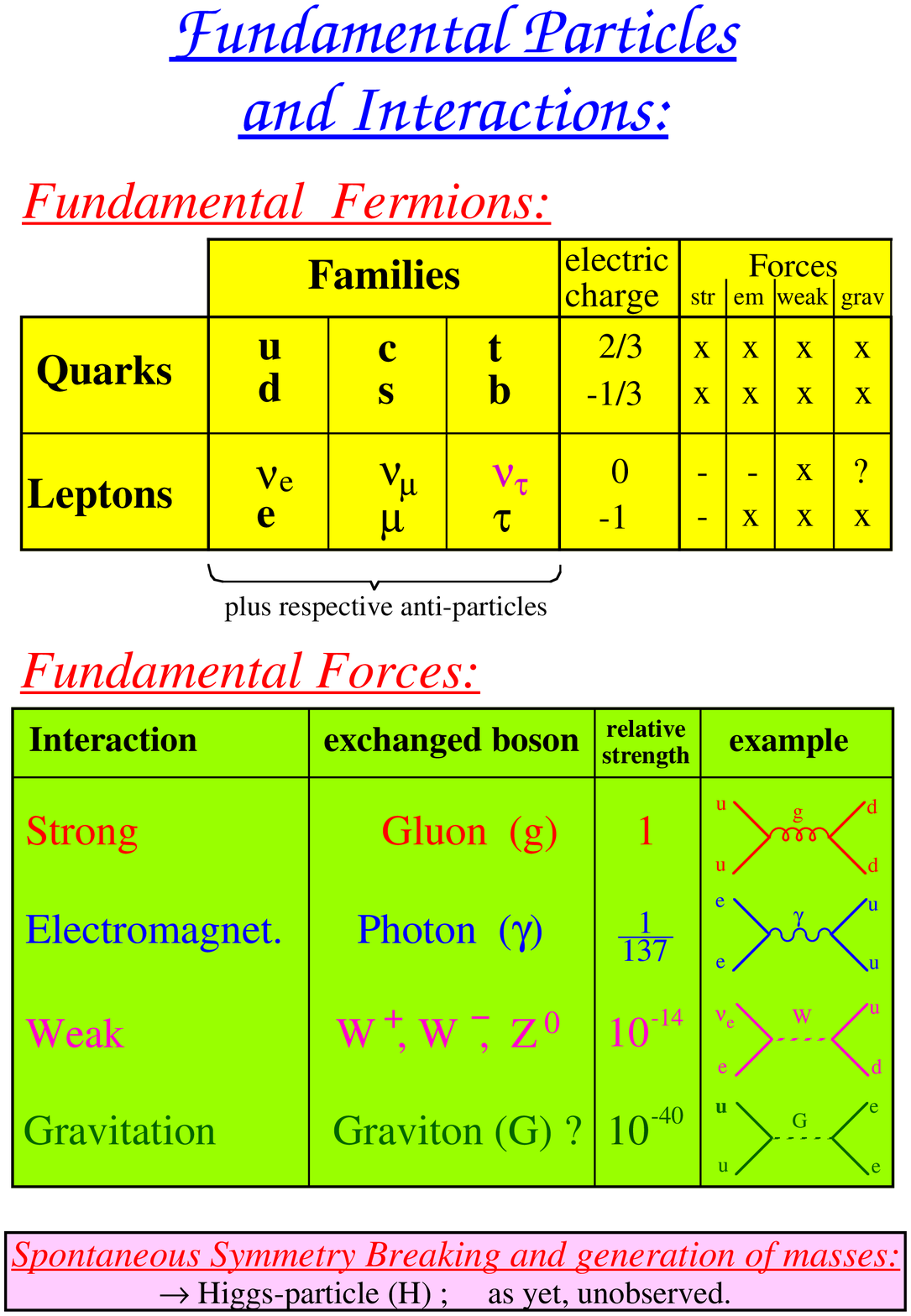,width=12.5cm}
\end{center}
\caption{\label{fig2}
}
\end{figure}

\begin{figure}[h,t]
\begin{center}
\epsfig{file=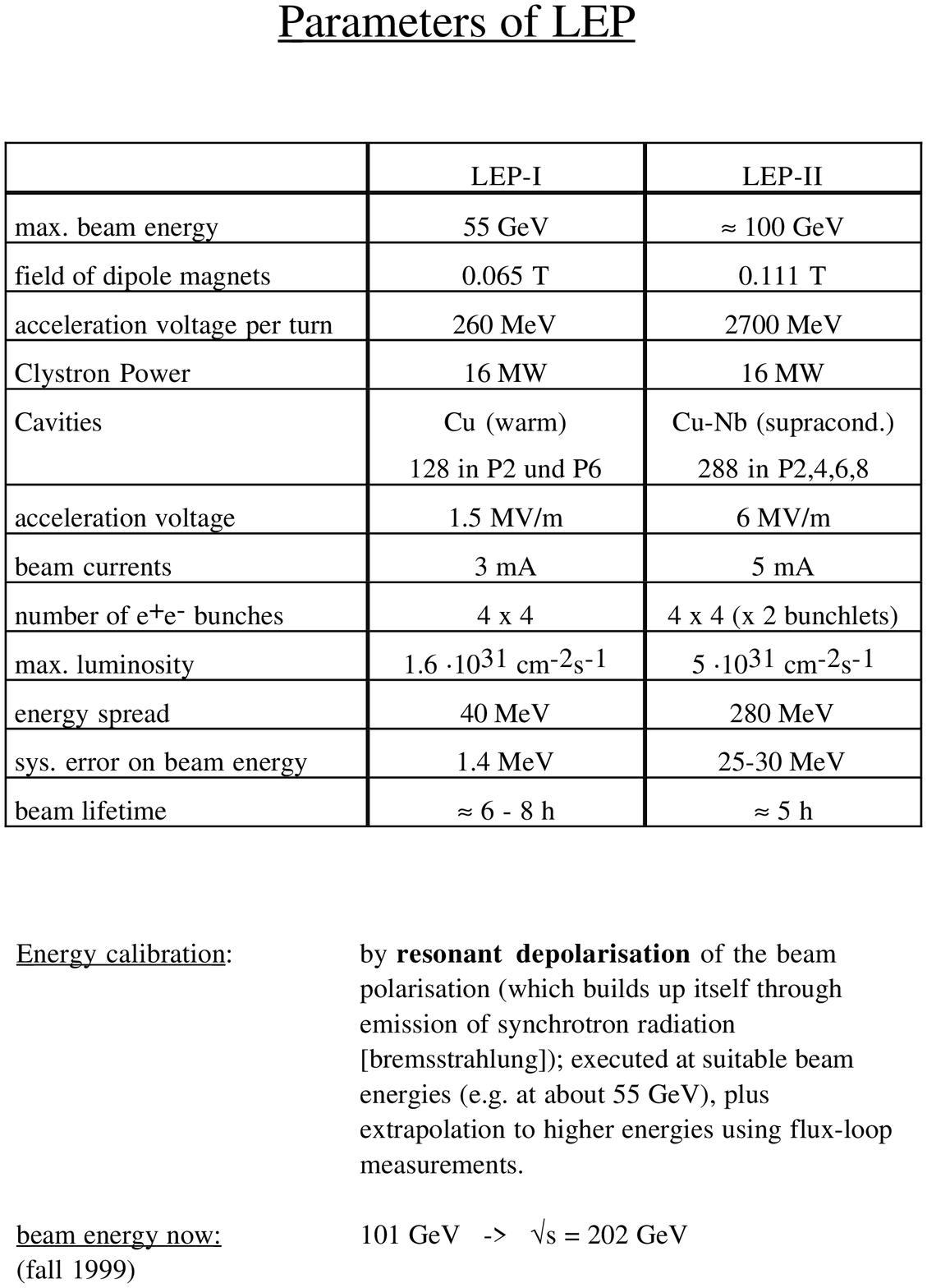,width=12.5cm}
\end{center}
\caption{\label{fig3}
}
\end{figure}

\begin{figure}[h,t]
\begin{center}
\epsfig{file=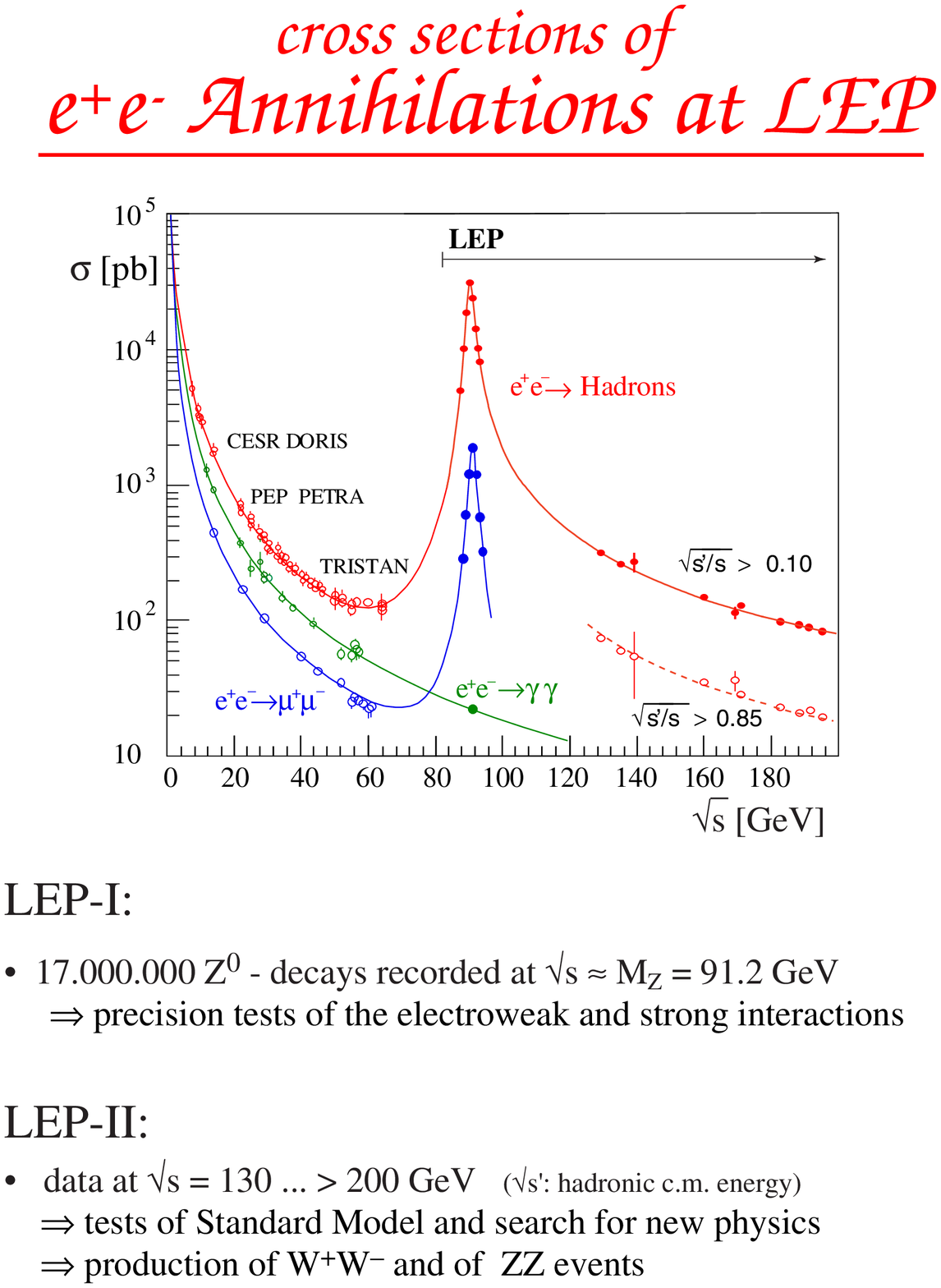,width=12.5cm}
\end{center}
\caption{\label{fig4}
}
\end{figure}

\begin{figure}[h,t]
\begin{center}
\epsfig{file=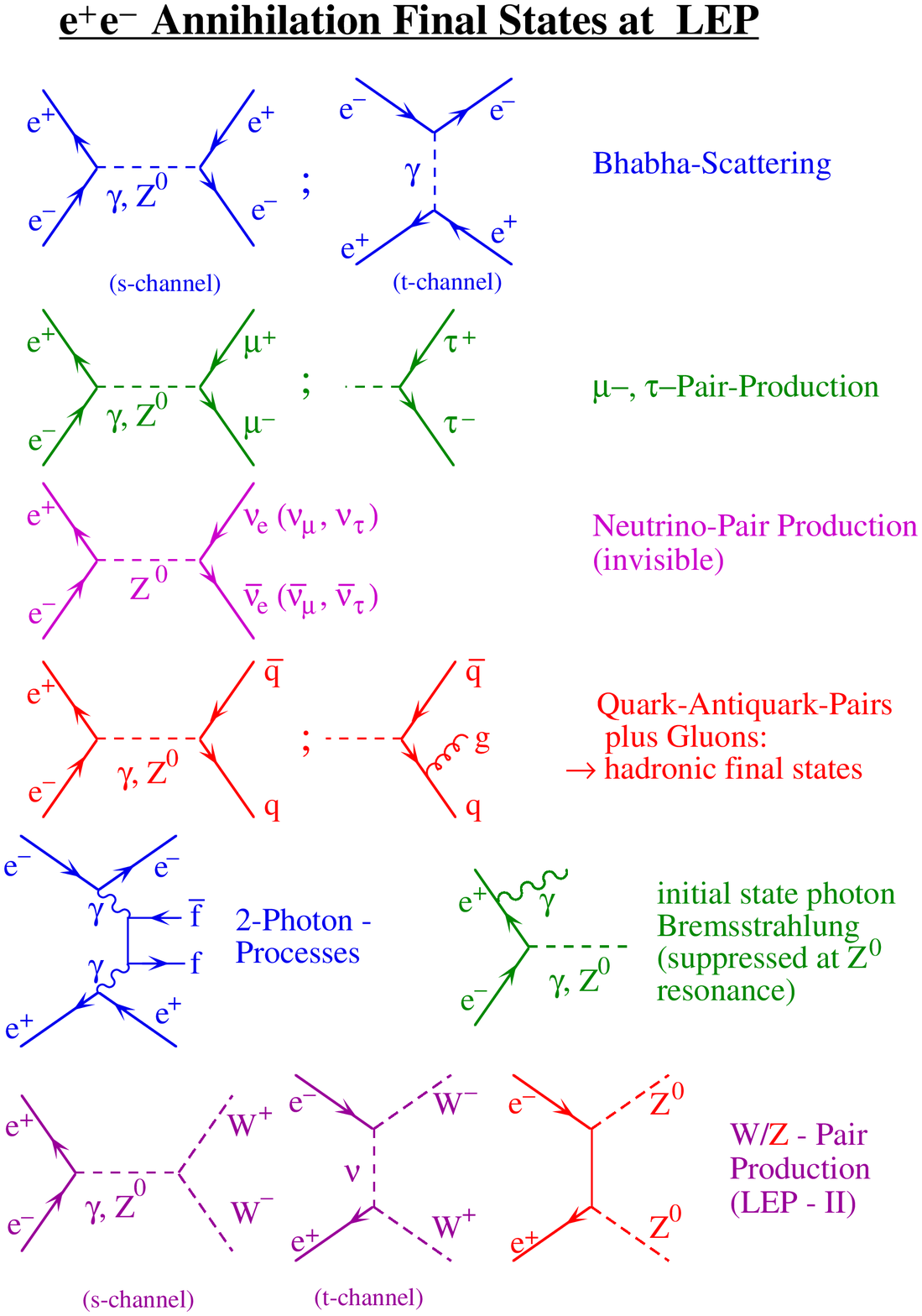,width=12.5cm}
\end{center}
\caption{\label{fig5}
}
\end{figure}

\begin{figure}[h,t]
\begin{center}
\epsfig{file=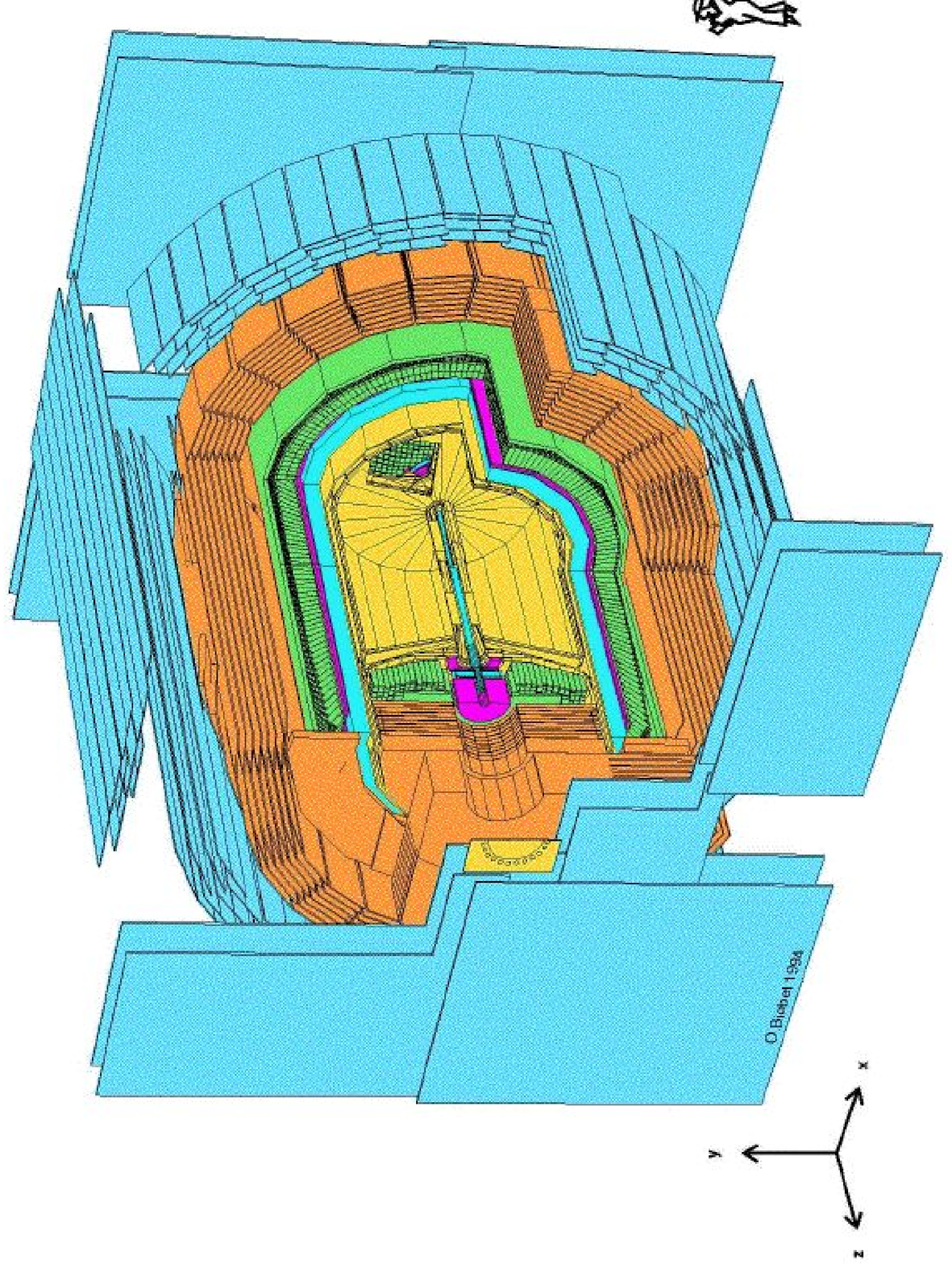,width=12.5cm}
\end{center}
\caption{\label{fig6}
}
\end{figure}

\begin{figure}[h,t]
\begin{center}
\epsfig{file=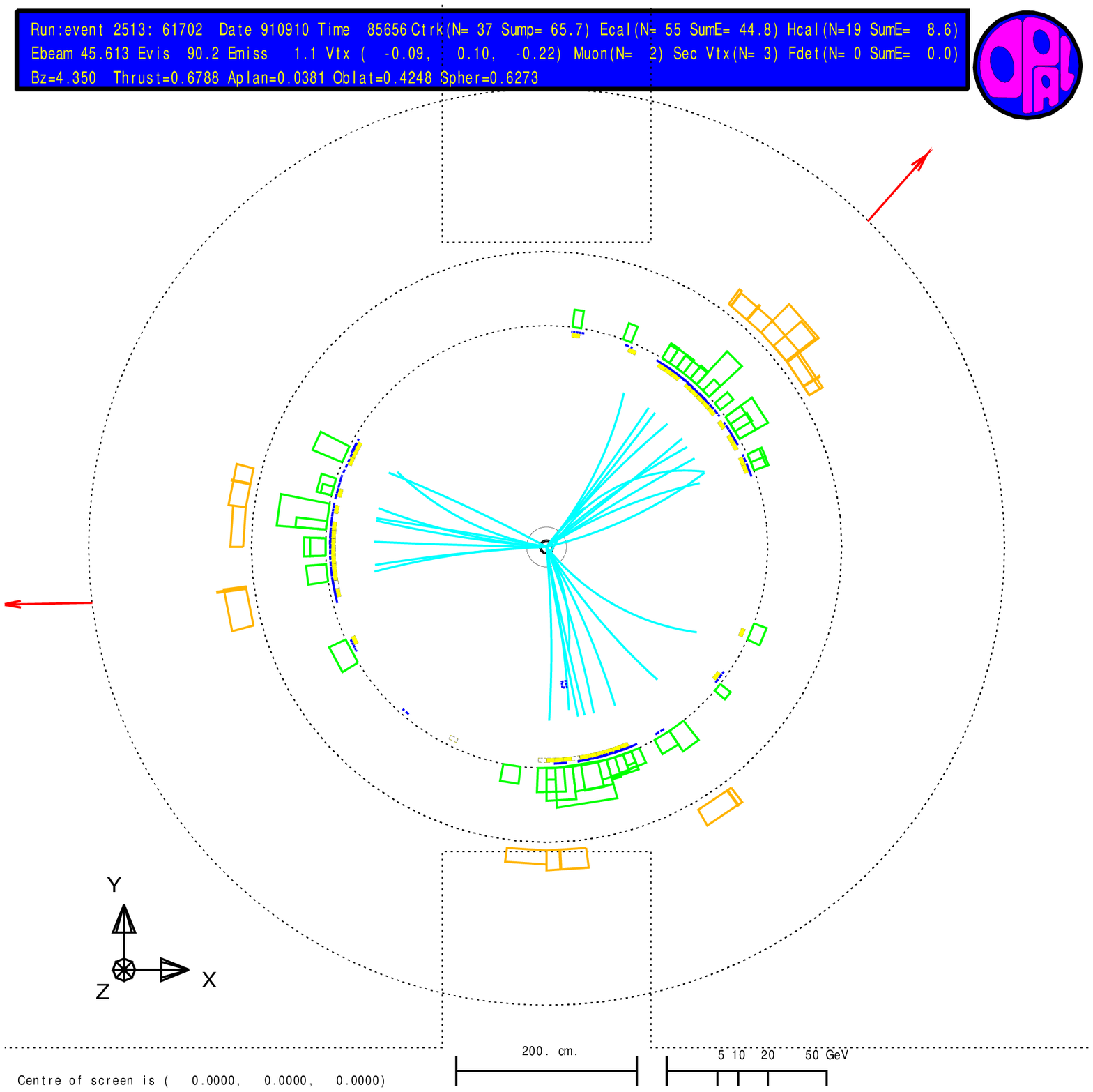,width=12.5cm}
\end{center}
\caption{\label{fig7}
}
\end{figure}

\begin{figure}[h,t]
\begin{center}
\epsfig{file=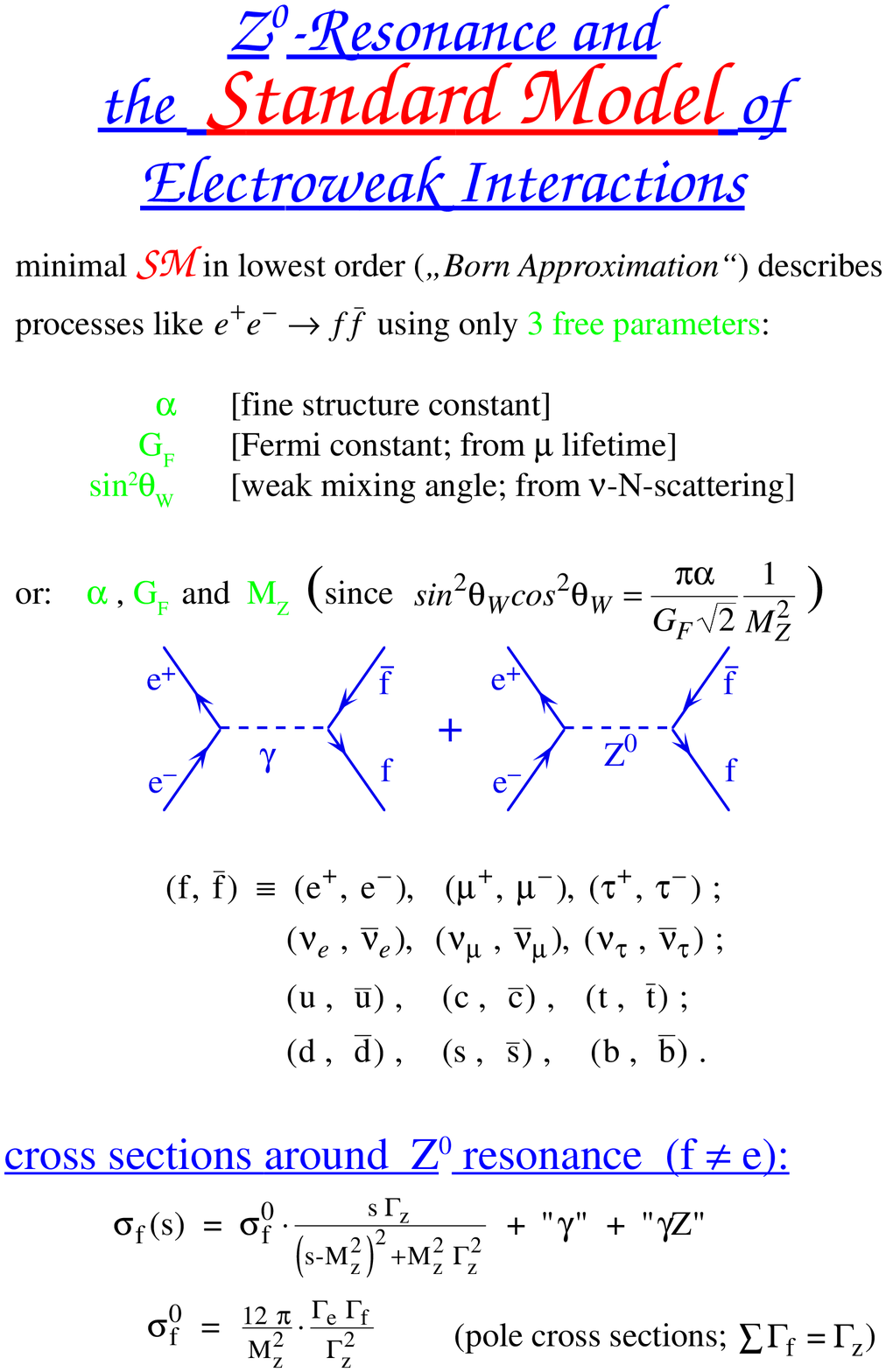,height=18.3cm}
\end{center}
\caption{\label{fig8}
}
\end{figure}

\begin{figure}[h,t]
\begin{center}
\epsfig{file=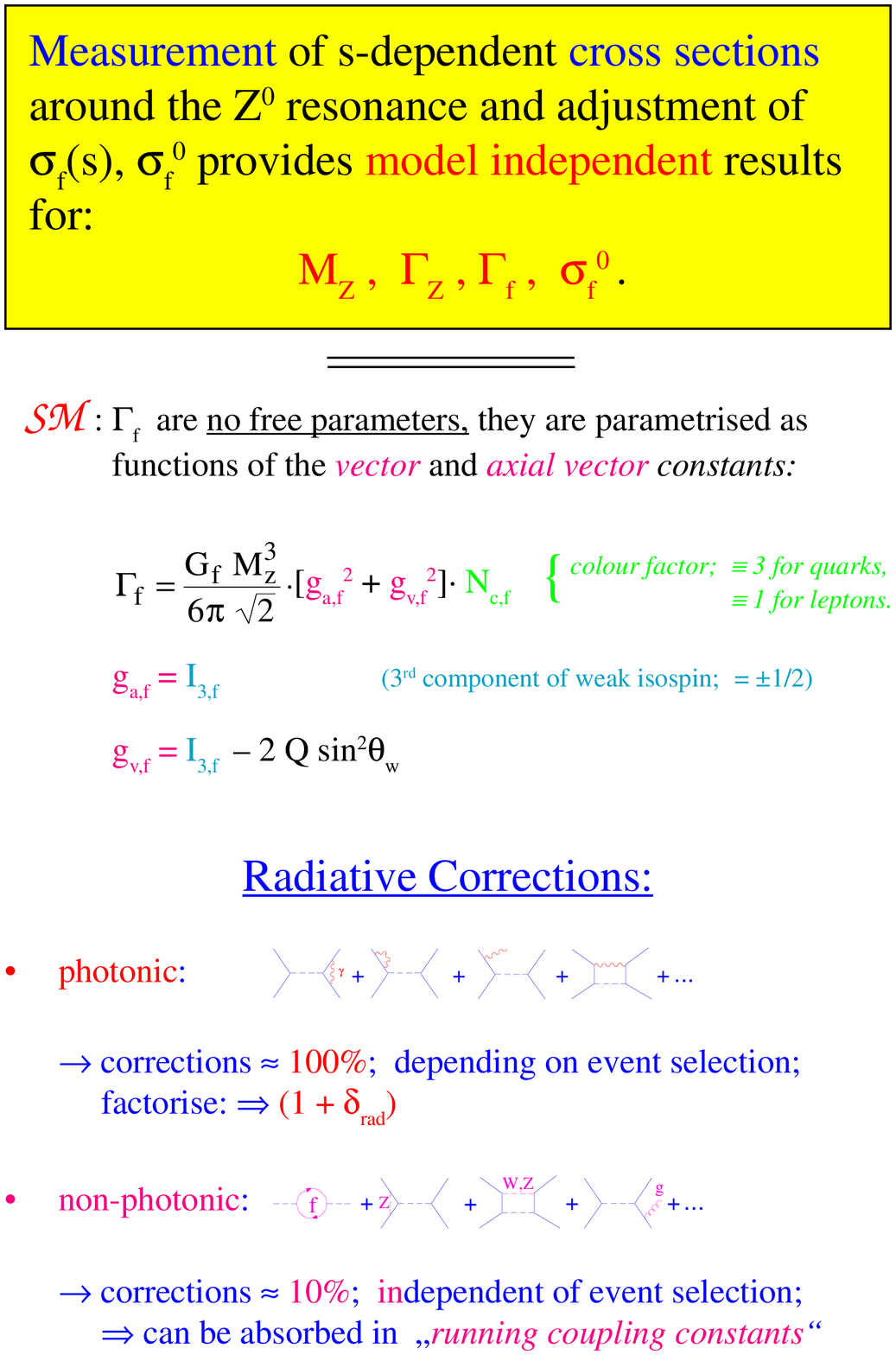,height=18.3cm}
\end{center}
\caption{\label{fig9}
}
\end{figure}

\begin{figure}[h,t]
\begin{center}
\epsfig{file=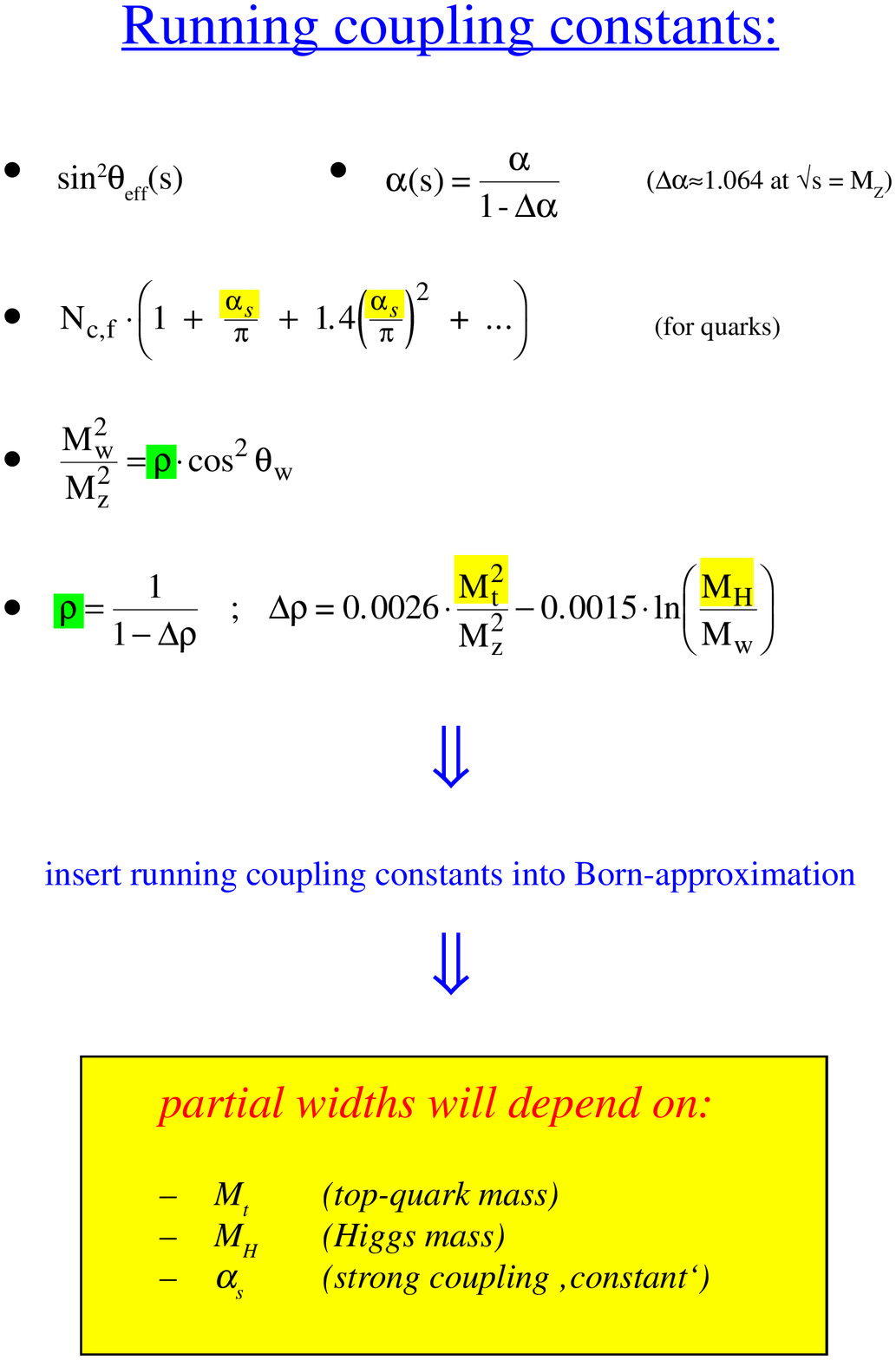,height=18.3cm}
\end{center}
\caption{\label{fig10}
}
\end{figure}

\begin{figure}[h,t]
\begin{center}
\epsfig{file=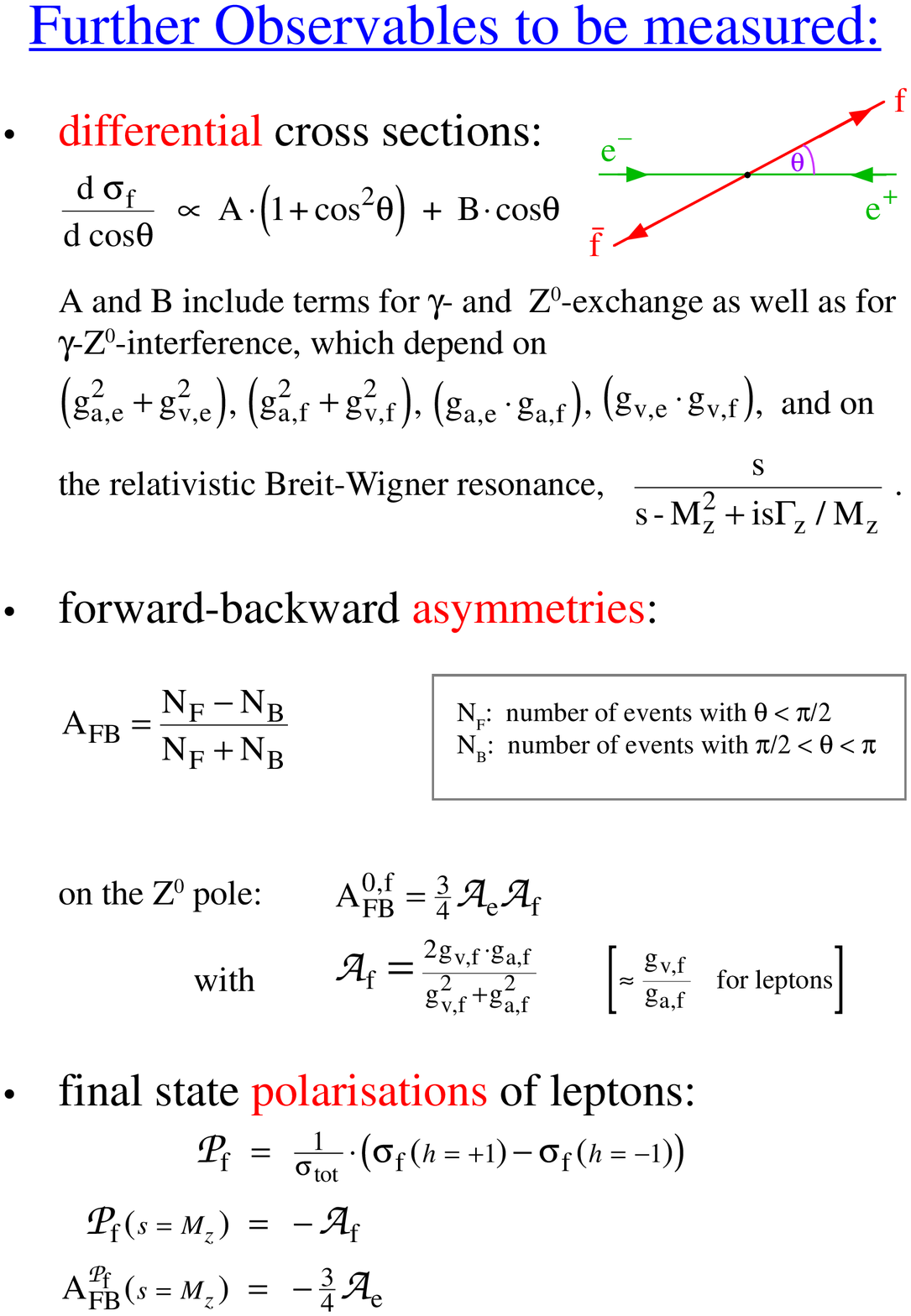,width=12.5cm}
\end{center}
\caption{\label{fig11}
}
\end{figure}

\begin{figure}[h,t]
\begin{center}
\epsfig{file=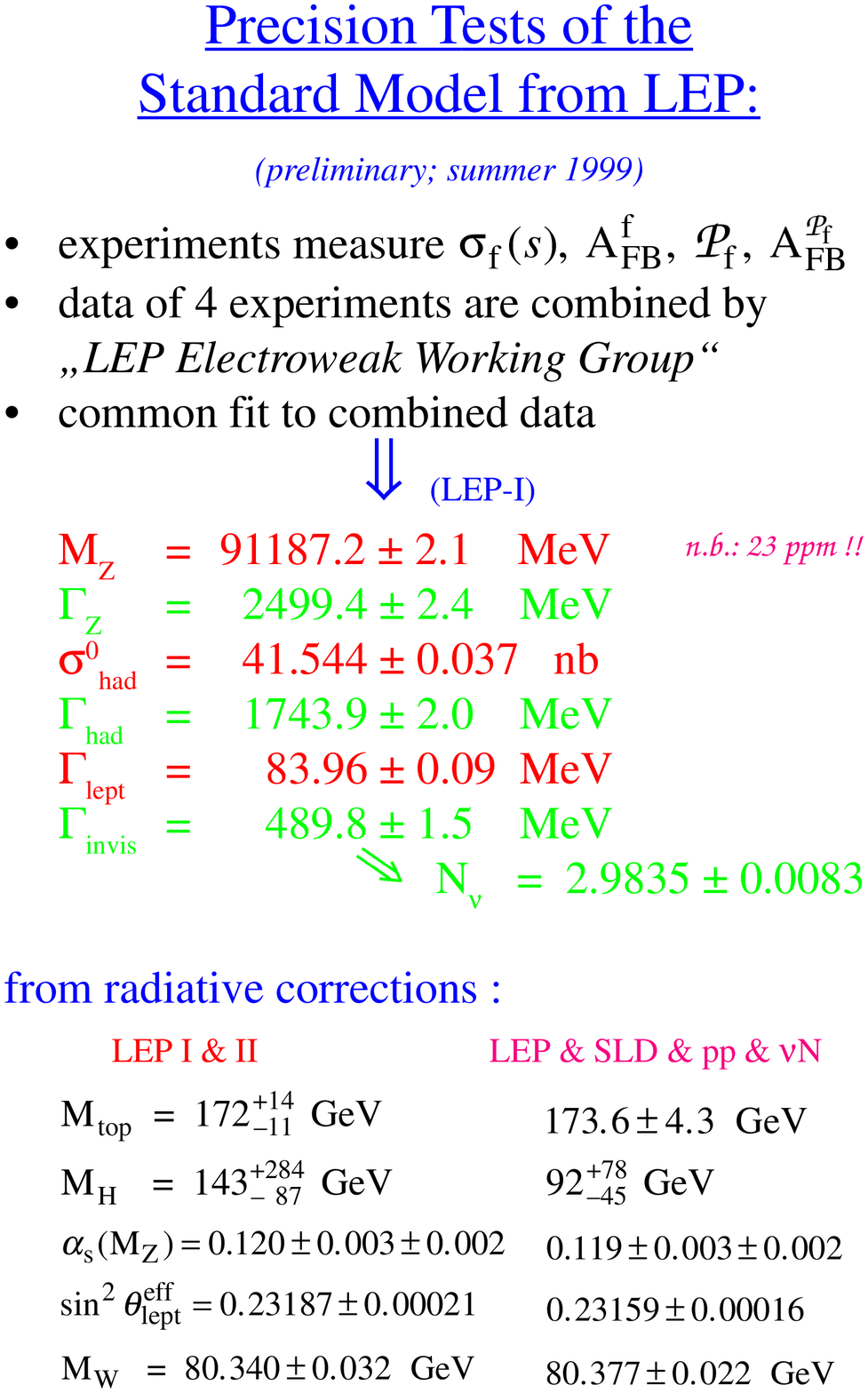,height=18.3cm}
\end{center}
\caption{\label{fig12}
}
\end{figure}

\begin{figure}[h,t]
\begin{center}
\epsfig{file=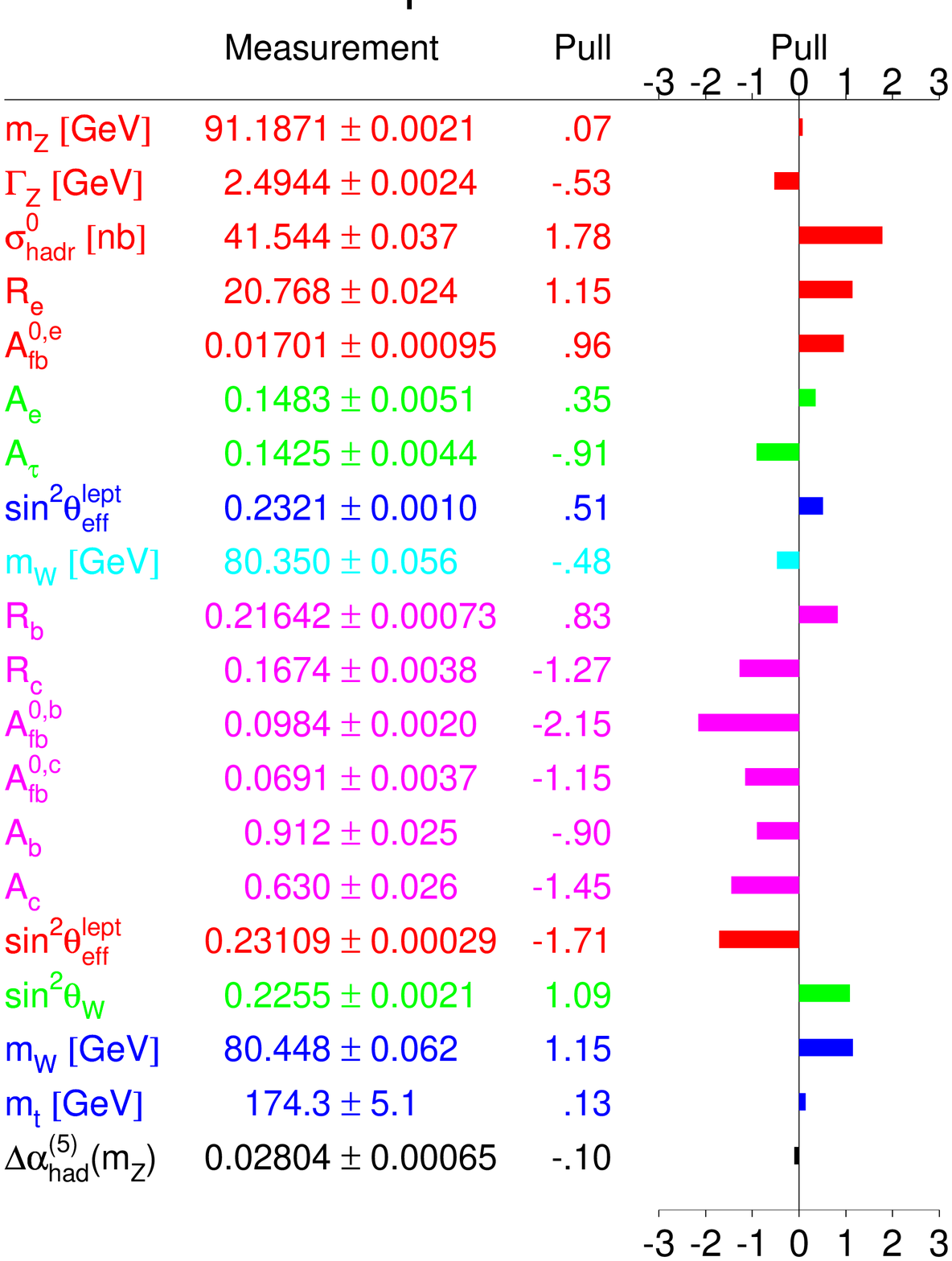,width=12.5cm}
\end{center}
\caption{\label{fig13}
}
\end{figure}

\begin{figure}[h,t]
\begin{center}
\epsfig{file=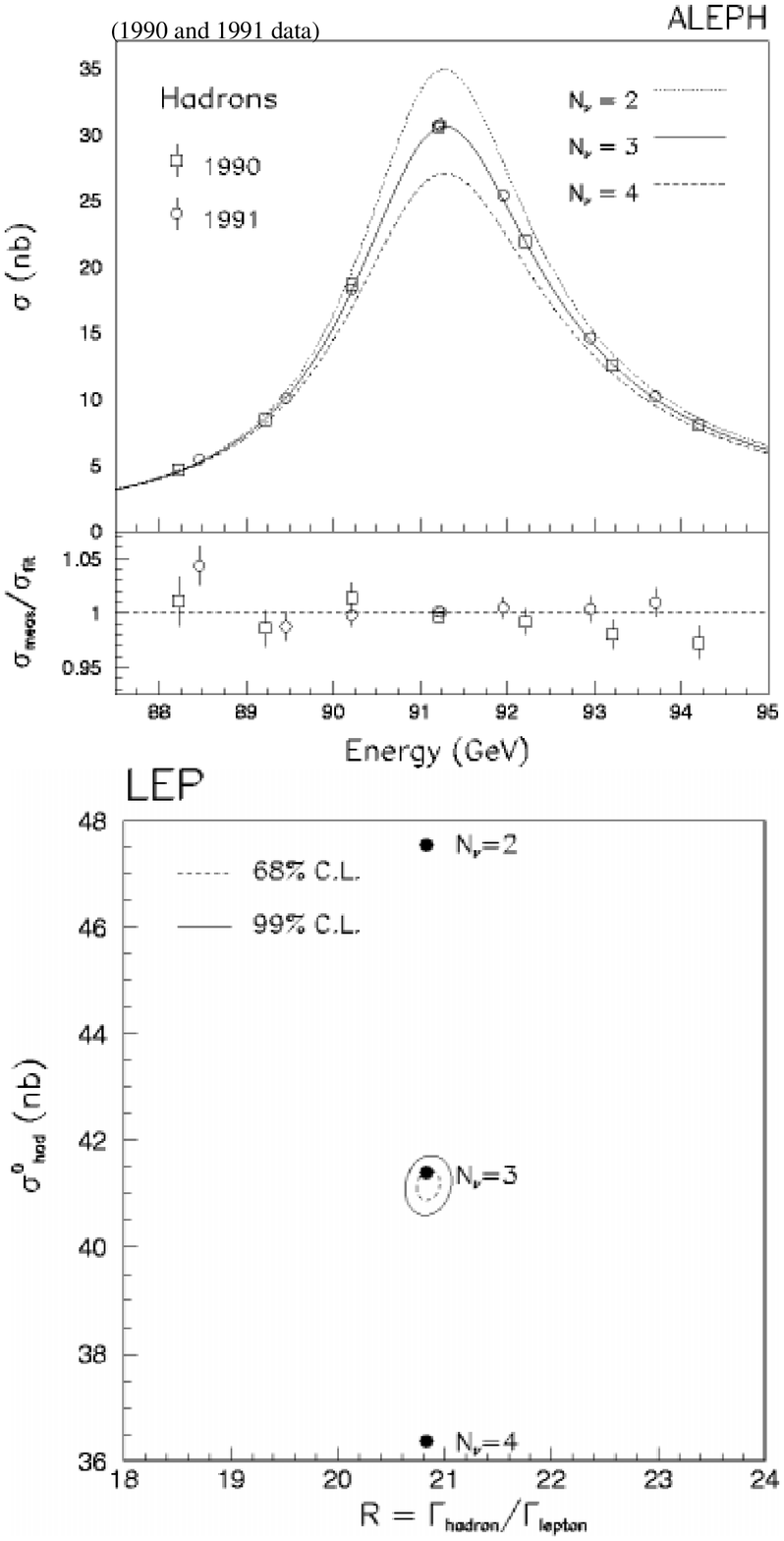,height=18.3cm}
\end{center}
\caption{\label{fig14}
}
\end{figure}

\begin{figure}[h,t]
\begin{center}
\epsfig{file=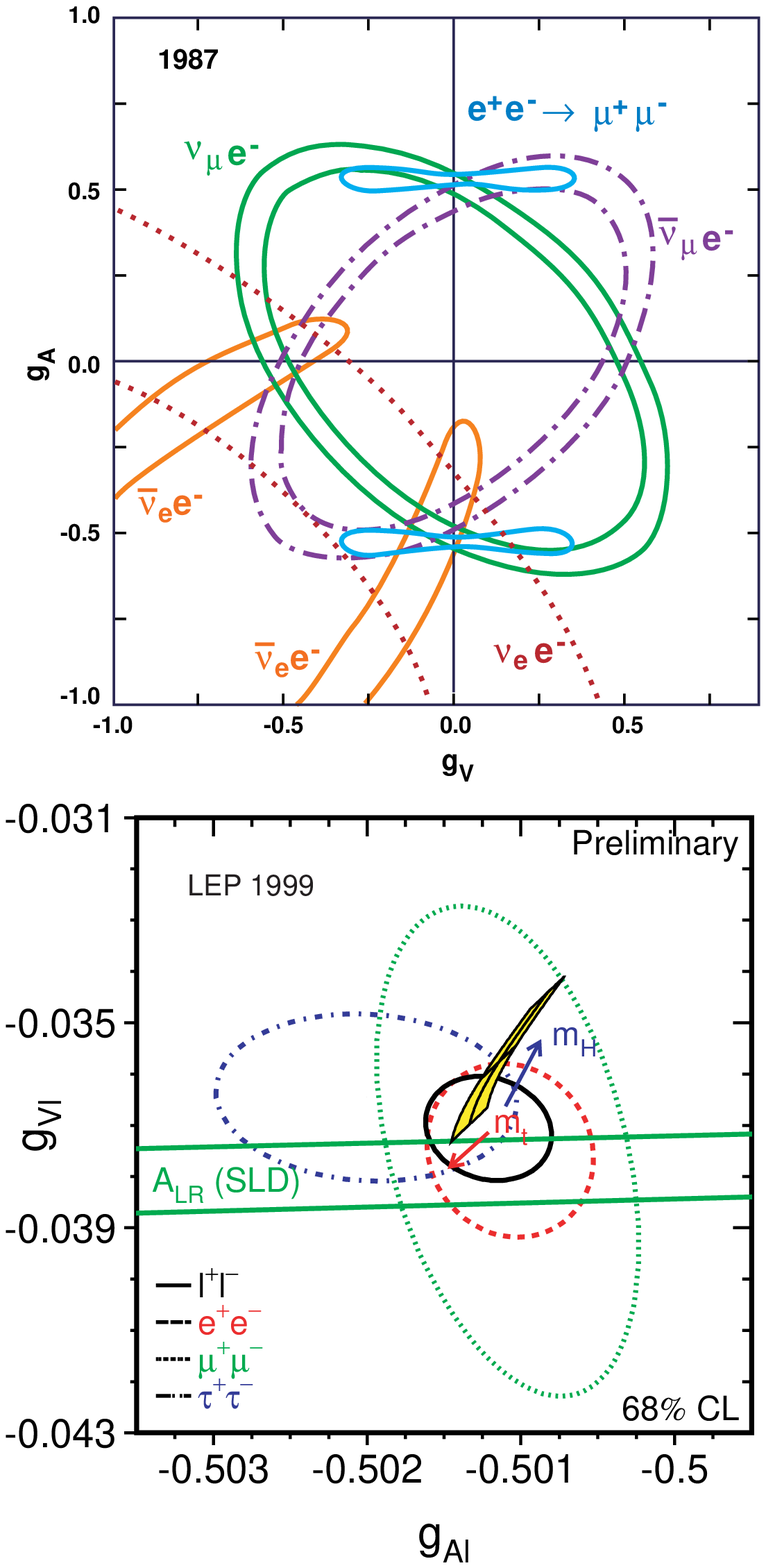,height=18.3cm}
\end{center}
\caption{\label{fig15}
}
\end{figure}

\begin{figure}[h,t]
\begin{center}
\epsfig{file=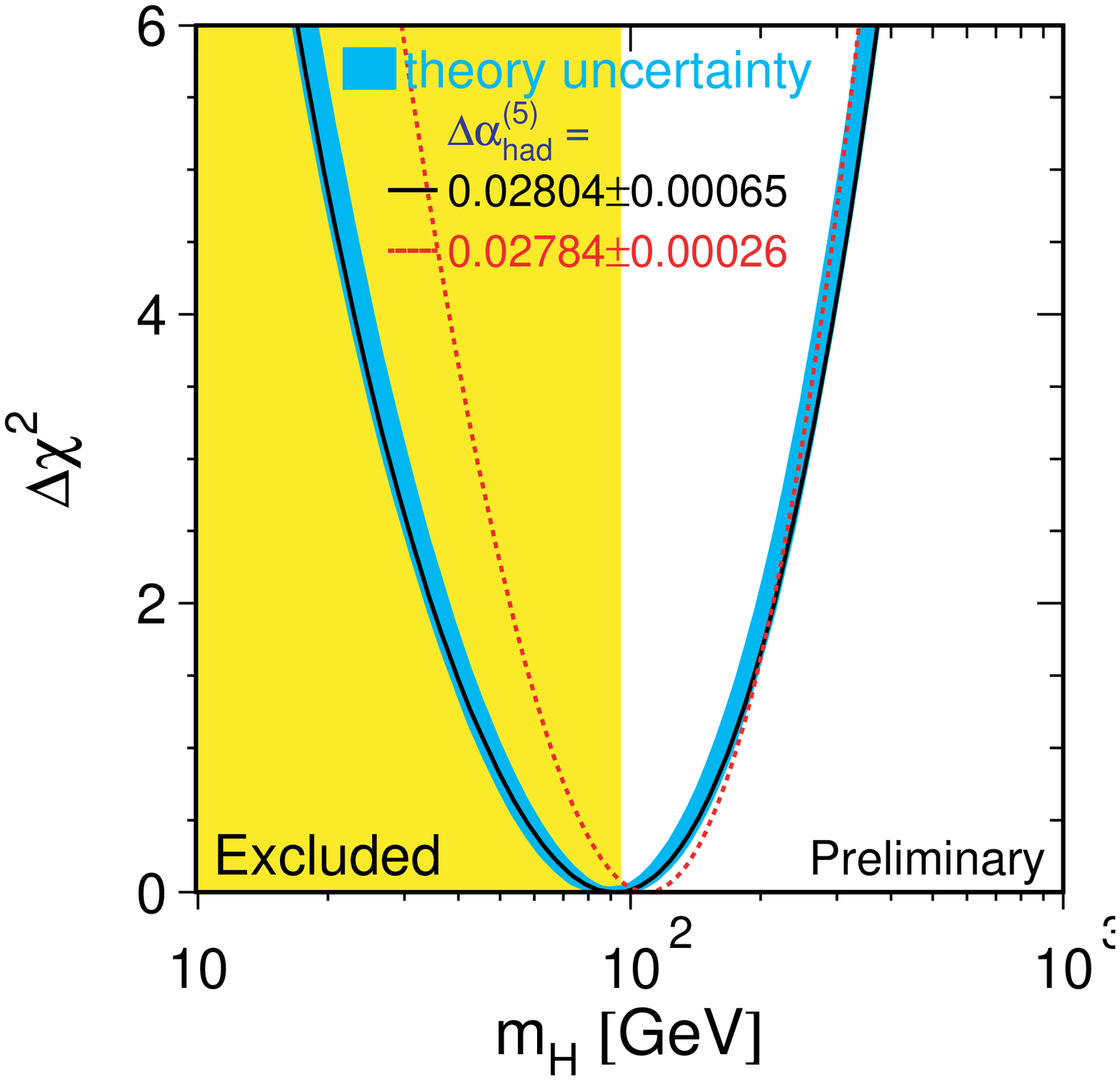,width=12.5cm}
\end{center}
\caption{\label{fig16}
}
\end{figure}

\begin{figure}[h,t]
\begin{center}
\epsfig{file=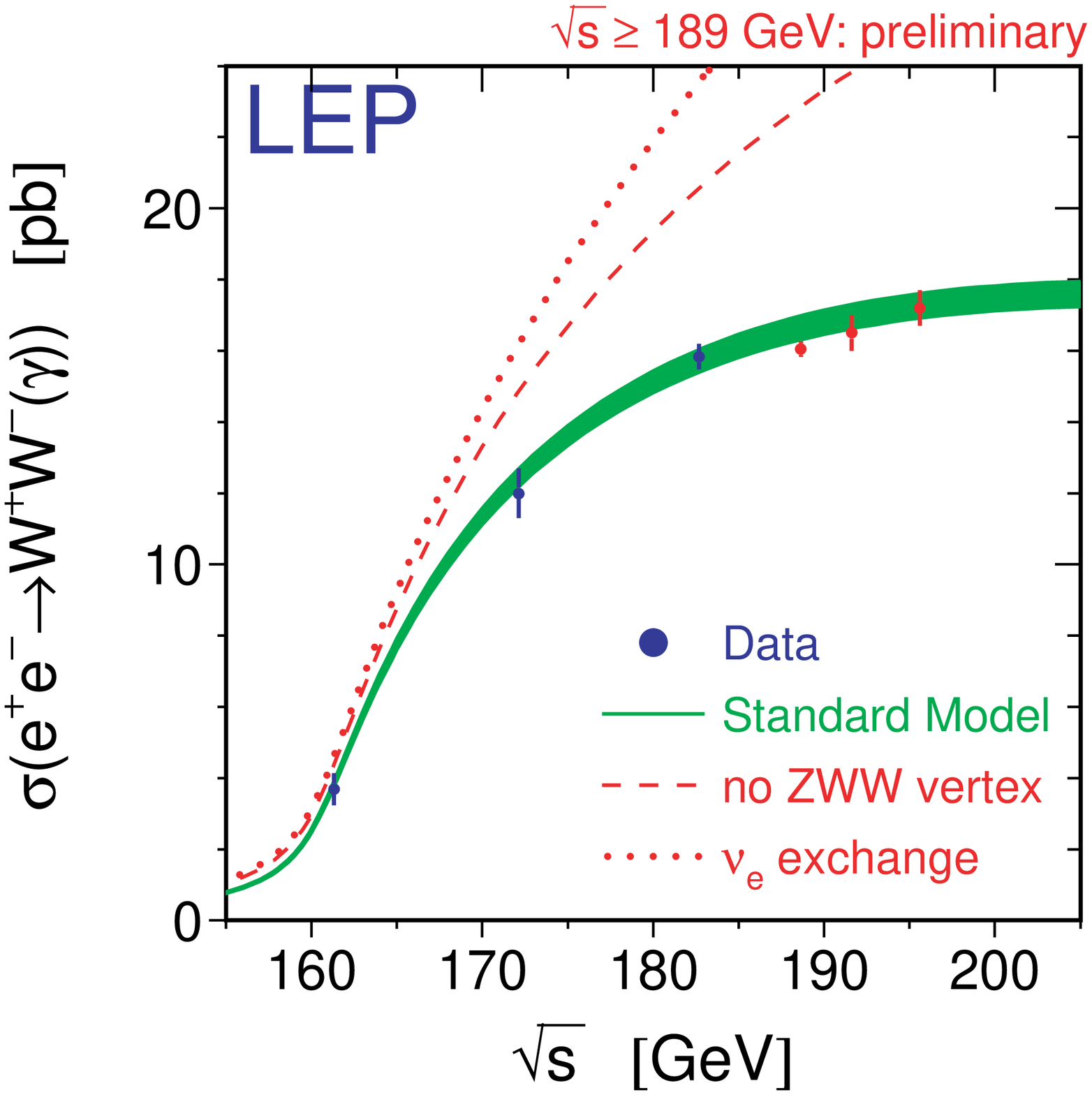,width=12.5cm}
\end{center}
\caption{\label{fig17}
}
\end{figure}
\clearpage

\begin{figure}[h,t]
\begin{center}
\epsfig{file=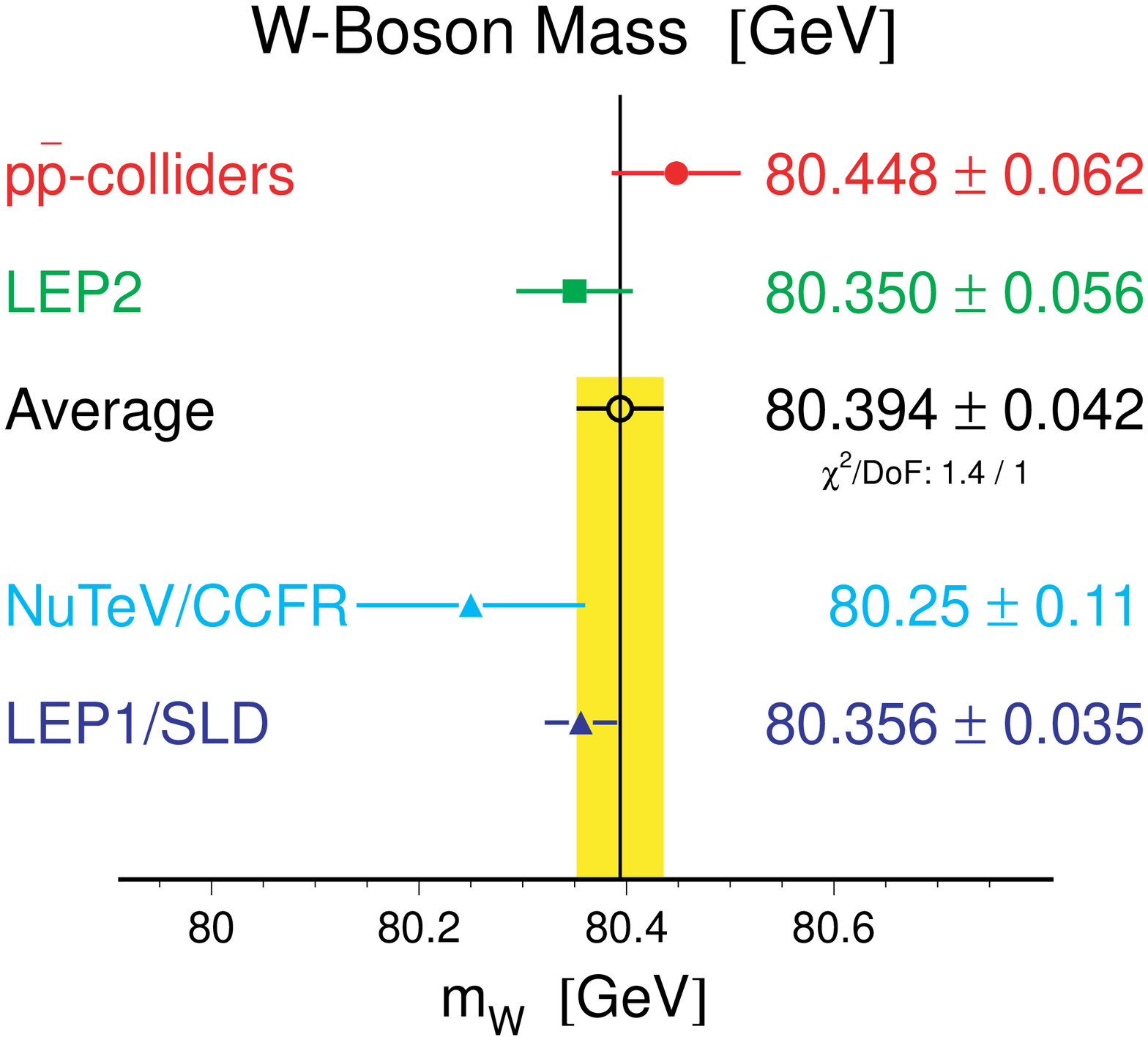,width=12.5cm}
\end{center}
\caption{\label{fig18}
}
\end{figure}

\begin{figure}[h,t]
\begin{center}
\epsfig{file=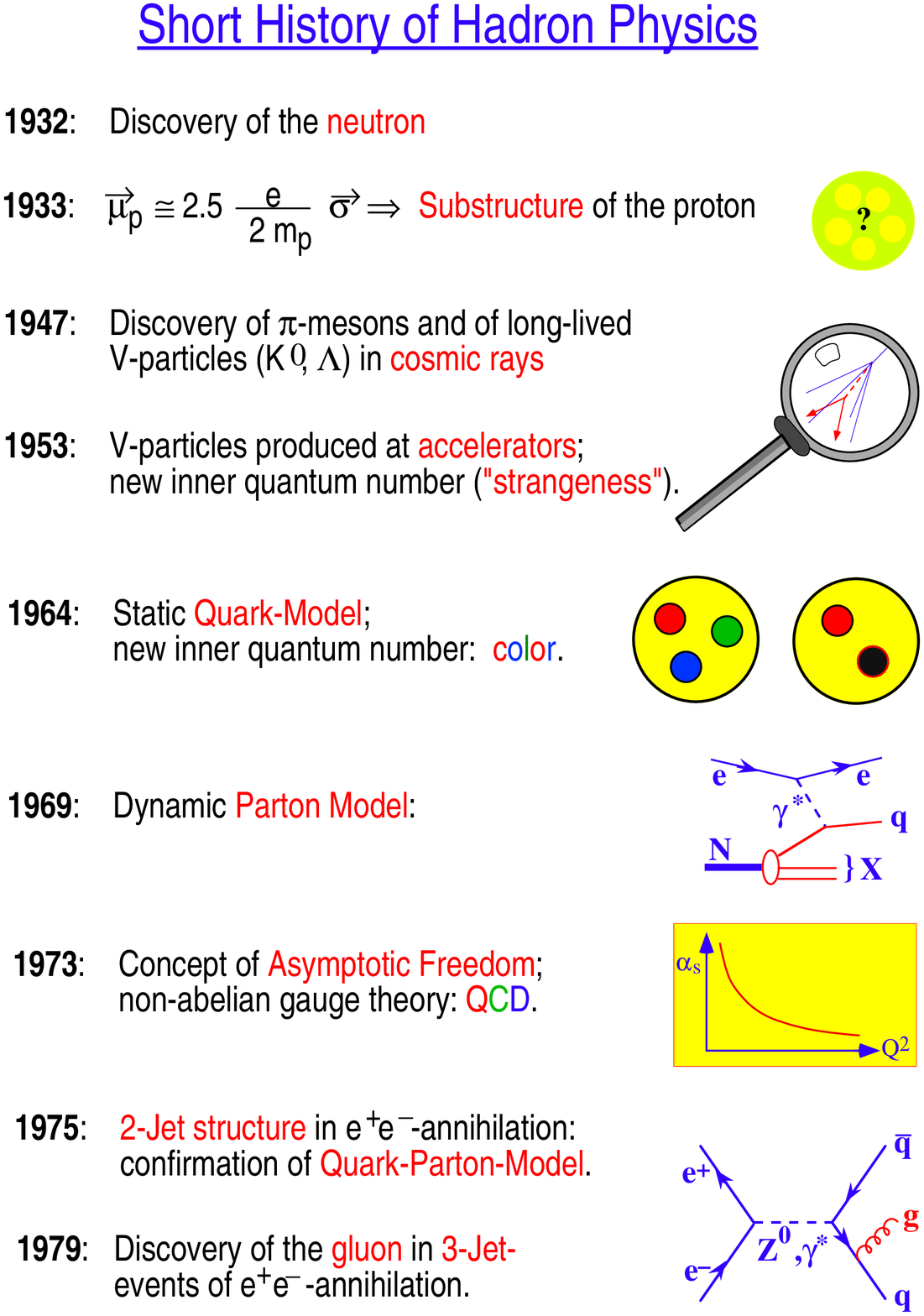,width=12.5cm}
\end{center}
\caption{\label{fig19}
}
\end{figure}

\begin{figure}[h,t]
\begin{center}
\epsfig{file=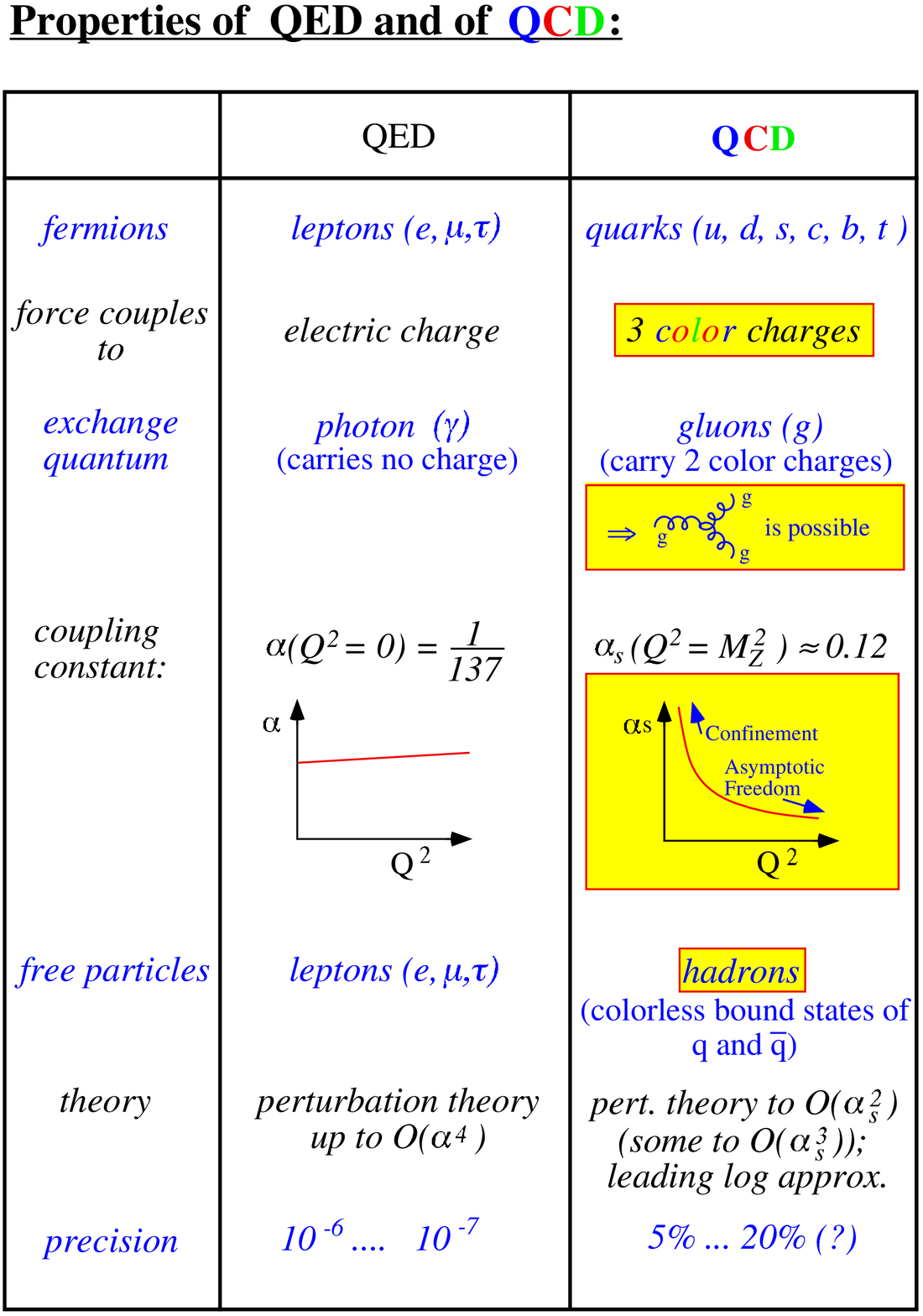,width=12.5cm}
\end{center}
\caption{\label{fig20}
}
\end{figure}

\begin{figure}[h,t]
\begin{center}
\epsfig{file=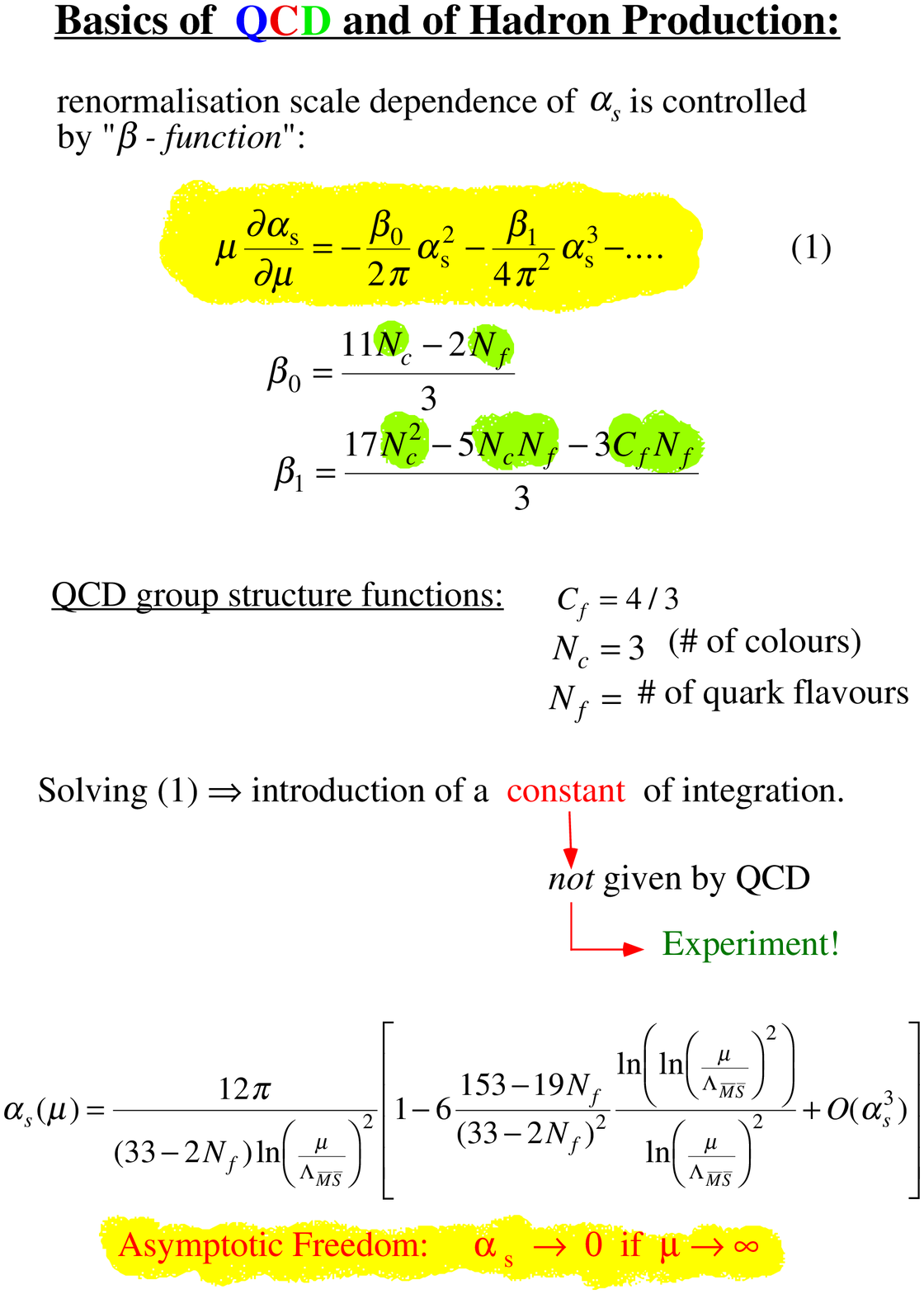,width=12.5cm}
\end{center}
\caption{\label{fig21}
}
\end{figure}

\begin{figure}[h,t]
\begin{center}
\epsfig{file=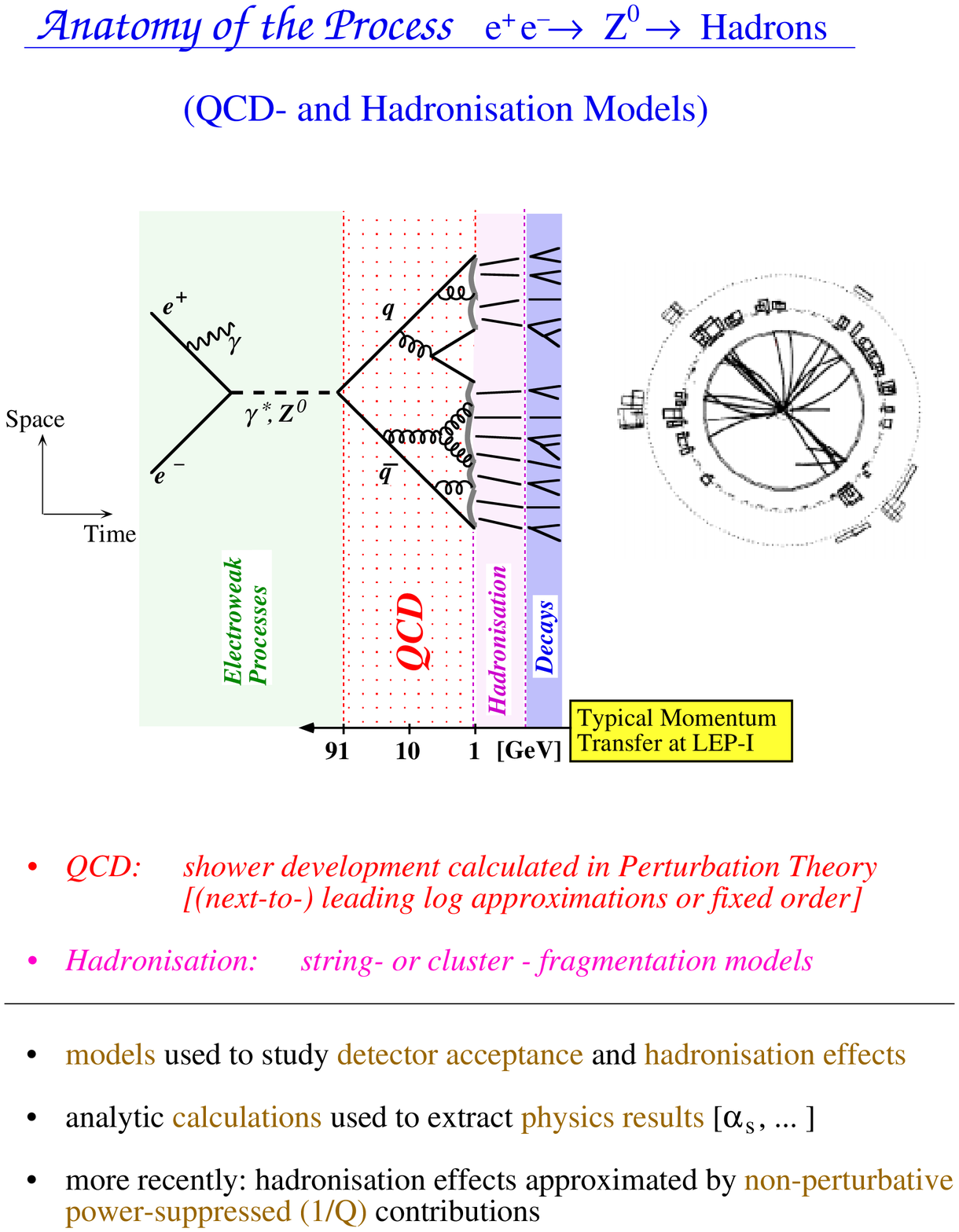,width=12.5cm}
\end{center}
\caption{\label{fig22}
}
\end{figure}

\begin{figure}[h,t]
\begin{center}
\epsfig{file=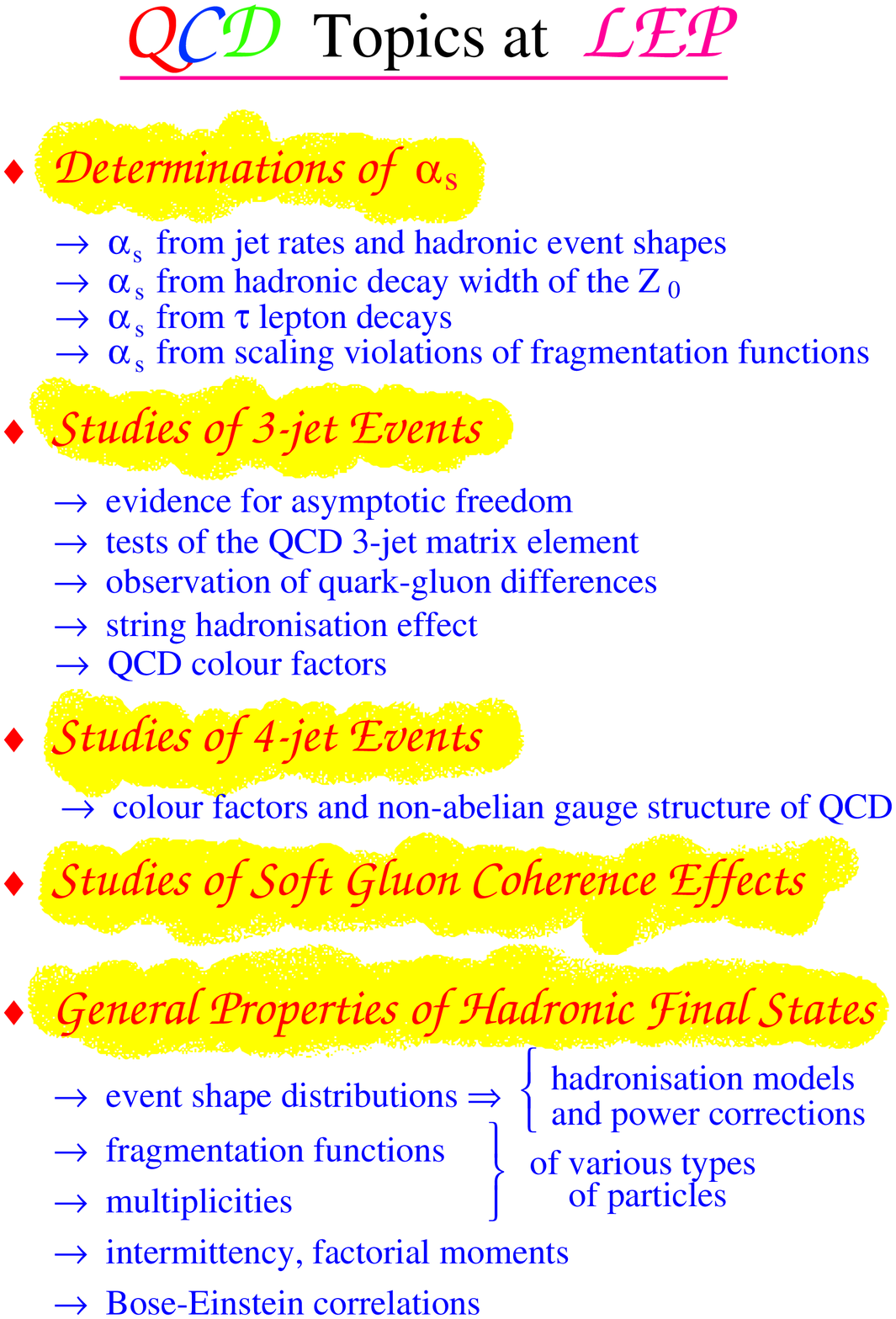,width=12.5cm}
\end{center}
\caption{\label{fig23}
}
\end{figure}

\begin{figure}[h,t]
\begin{center}
\epsfig{file=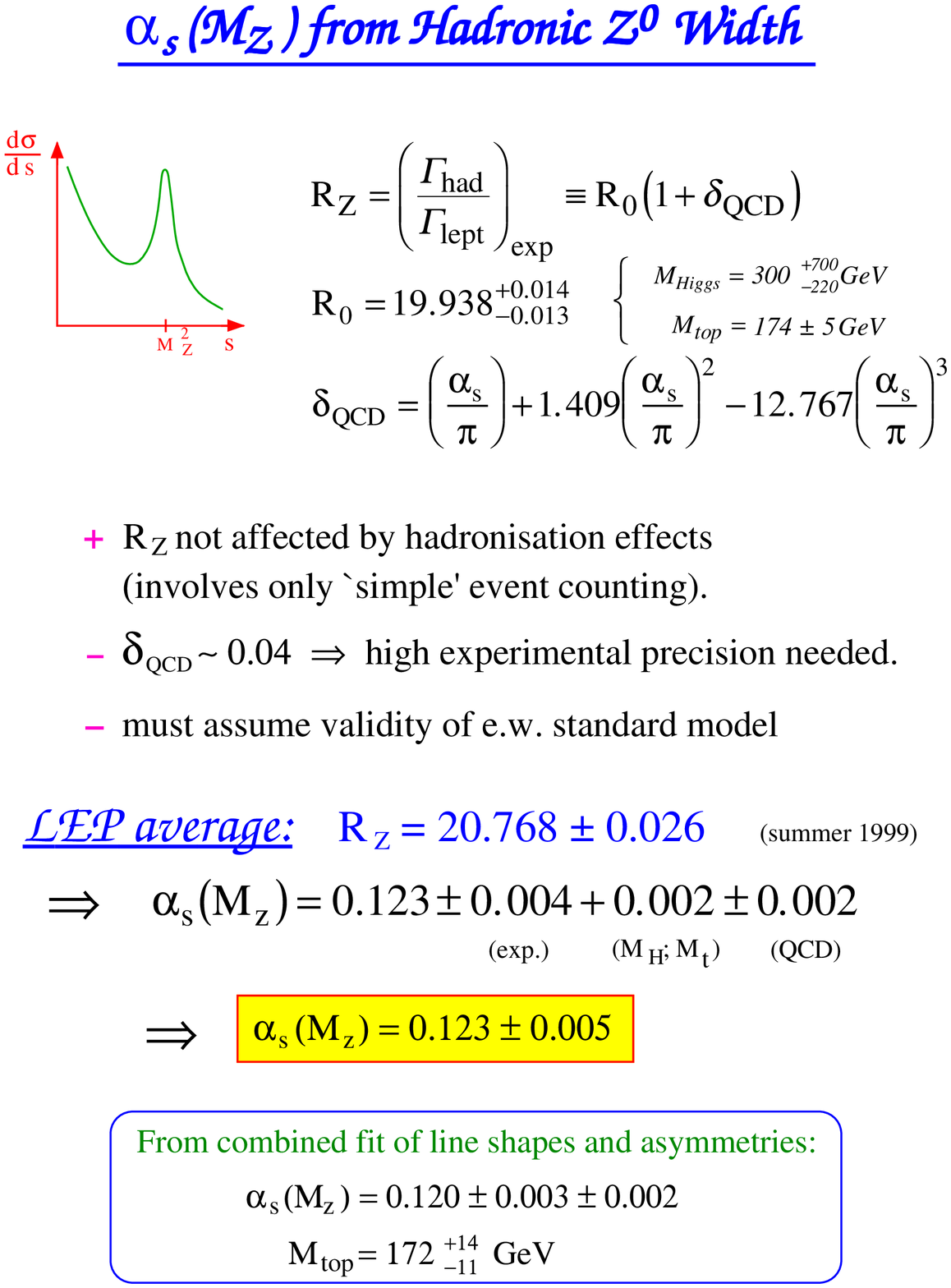,width=12.5cm}
\end{center}
\caption{\label{fig24}
}
\end{figure}

\begin{figure}[h,t]
\begin{center}
\epsfig{file=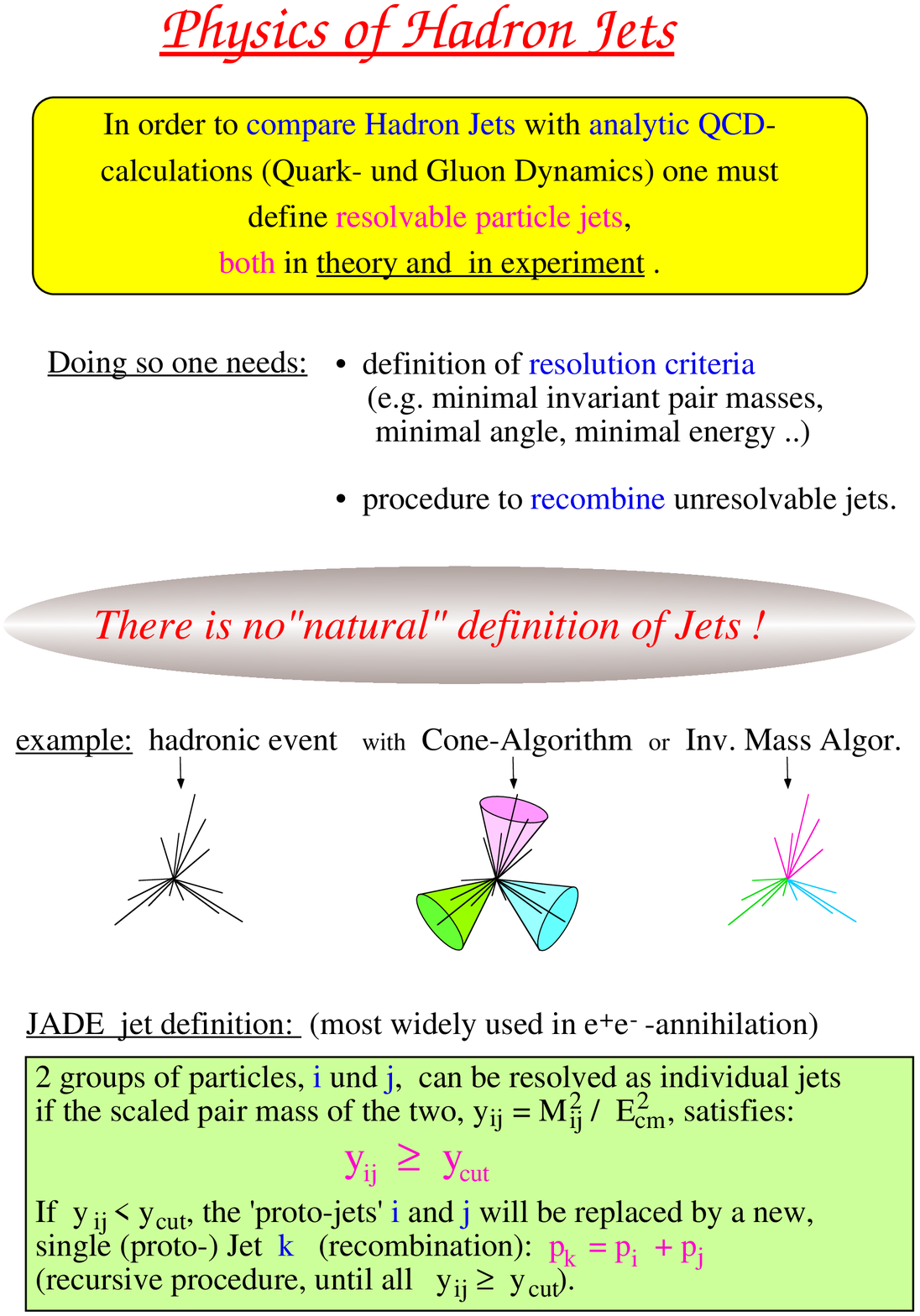,width=12.5cm}
\end{center}
\caption{\label{fig25}
}
\end{figure}
\clearpage

%
%
\renewcommand{\arraystretch}{1.2}
\begin{center}
{\huge JADE-type Jet Cluster Algorithms}
\end{center}
\begin{table}[h]
\begin{tabular}{|c||c|p{1.20in}|p{1.70in}|}
   \hline
     Algorithm& Resolution $y_{ij}$ & Recombination & Remarks  \\
   \hline \hline
JADE & \large{${2 E_i E_j (1 - \cos \theta_{ij})}\over{s}$} &
$p_k = p_i + p_j$ &
  conserves $\sum E$, $\sum \vec{p}$; \\ 
    & $[\equiv y_{ij}^J]$ & & does not exponentiate\\  
\hline
E  & \large{${(p_i + p_j)^2}\over{s}$} & $p_k = p_i + p_j$ &
Lorentz invariant \\
\hline
E0 & \large{${(p_i + p_j)^2}\over{s}$} & $E_k = E_i + E_j$; &
conserves $\sum E$, but\\
     & & $\vec{p}_k = \frac{E_k}
  {|\vec{p}_i + \vec{p}_j|} (\vec{p}_i + \vec{p}_j)$ &
  violates $\sum \vec{p}$ \\ \hline
p & \large{${(p_i + p_j)^2}\over{s}$} &
$\vec{p}_k = \vec{p}_i + \vec{p}_j$; &
 conserves $\sum \vec{p}$, but\\
  & & $E_k = |\vec{p}_k|$ & violates $\sum E$ \\
\hline
p0 & \large{${(p_i + p_j)^2}\over{s}$}&
$\vec{p}_k = \vec{p}_i + \vec{p}_j$; &
  as p-scheme; $s \equiv \sum E$ up- \\
  & & $E_k = |\vec{p}_k|$ & dated after each recomb. \\
 \hline \hline
{\bf D}urham & \large{${2\cdot{\rm min}(E_i^2, E_j^2)\cdot
(1 - \cos \theta_{ij})}\over{s}$}
& $p_k = p_i + p_j$ &
  conserves $\sum E$, $\sum \vec{p}$;\\
    & $[\equiv y_{ij}^D]$ & & avoids exp.
  problems\\ \hline 
{\bf C}ambridge &  ${2\cdot(1 - \cos \theta_{ij})}$;\ \ \ {\it soft}
& $p_k = p_i + p_j$ &
  conserves $\sum E$, $\sum \vec{p}$;\\
    & {\it freezing if} $y_{ij}^D > y_{cut}^{D}$ & & avoids exp.
  problems\\ \hline 
\hline
{\bf G}eneva & \large{${8 E_i E_j (1 - \cos \theta_{ij})}\over{9(E_i + E_j)^2}$}
& $p_k = p_i + p_j$ &
  conserves $\sum E$, $\sum \vec{p}$;\\
  &  & & avoids exp. problems\\
  \hline \hline
LUCLUS&   \large{$\frac{2 |\vec{p_i}| \cdot |\vec{p_j}| \cdot
\sin(\theta_{ij}/2)}
  {|\vec{p_i}| + |\vec{p_j}|}$ }  &
  $p_k = p_i + p_j$ & conserves $\sum E$, $\sum \vec{p}$; \\
& & & uncalculable in pert. th. \\
\hline
   \end{tabular}
\caption{\label{fig26}}
\end{table}
\clearpage

\begin{figure}[h,t]
\begin{center}
\epsfig{file=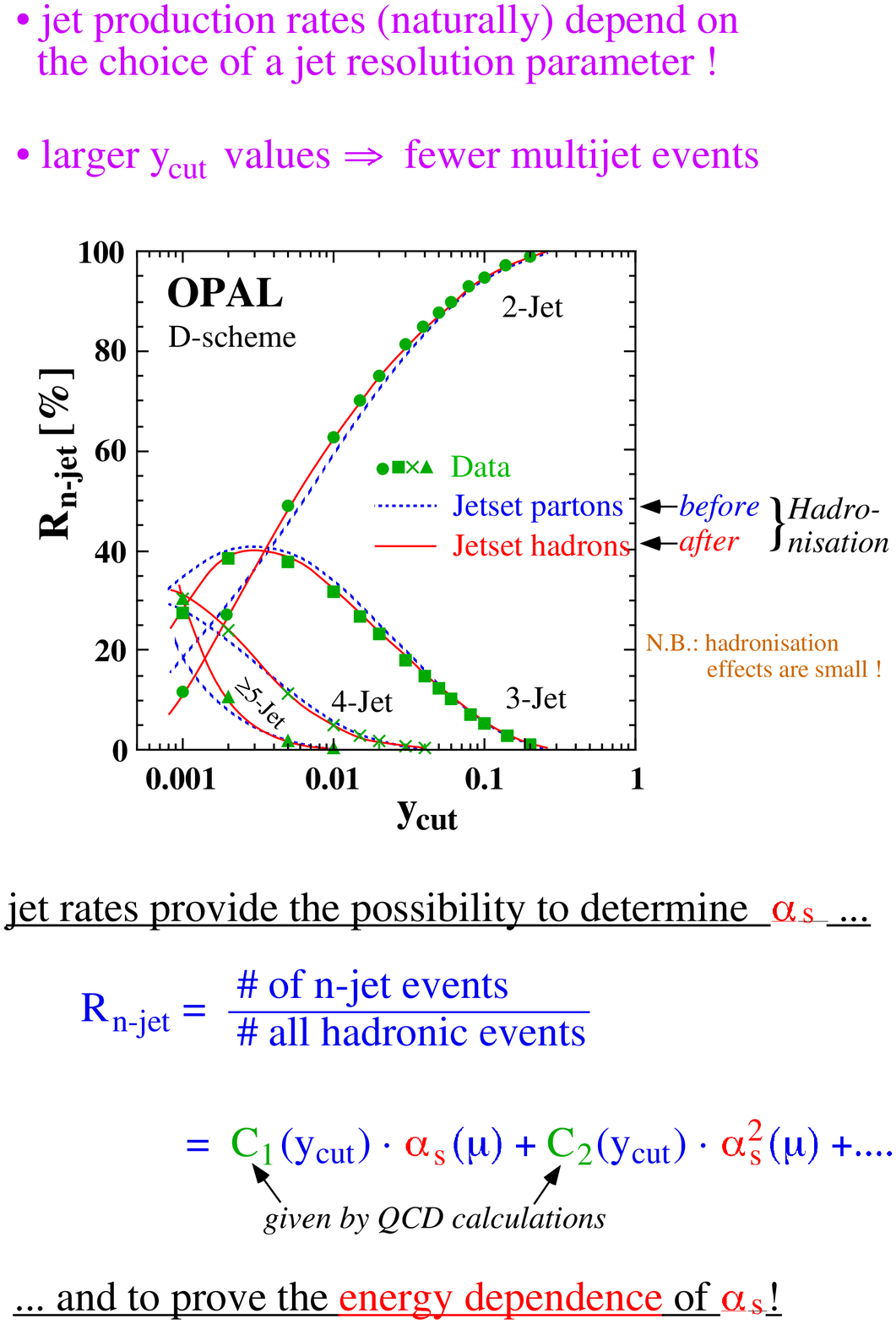,width=12.5cm}
\end{center}
\caption{\label{fig27}
}
\end{figure}

\begin{figure}[h,t]
\begin{center}
\epsfig{file=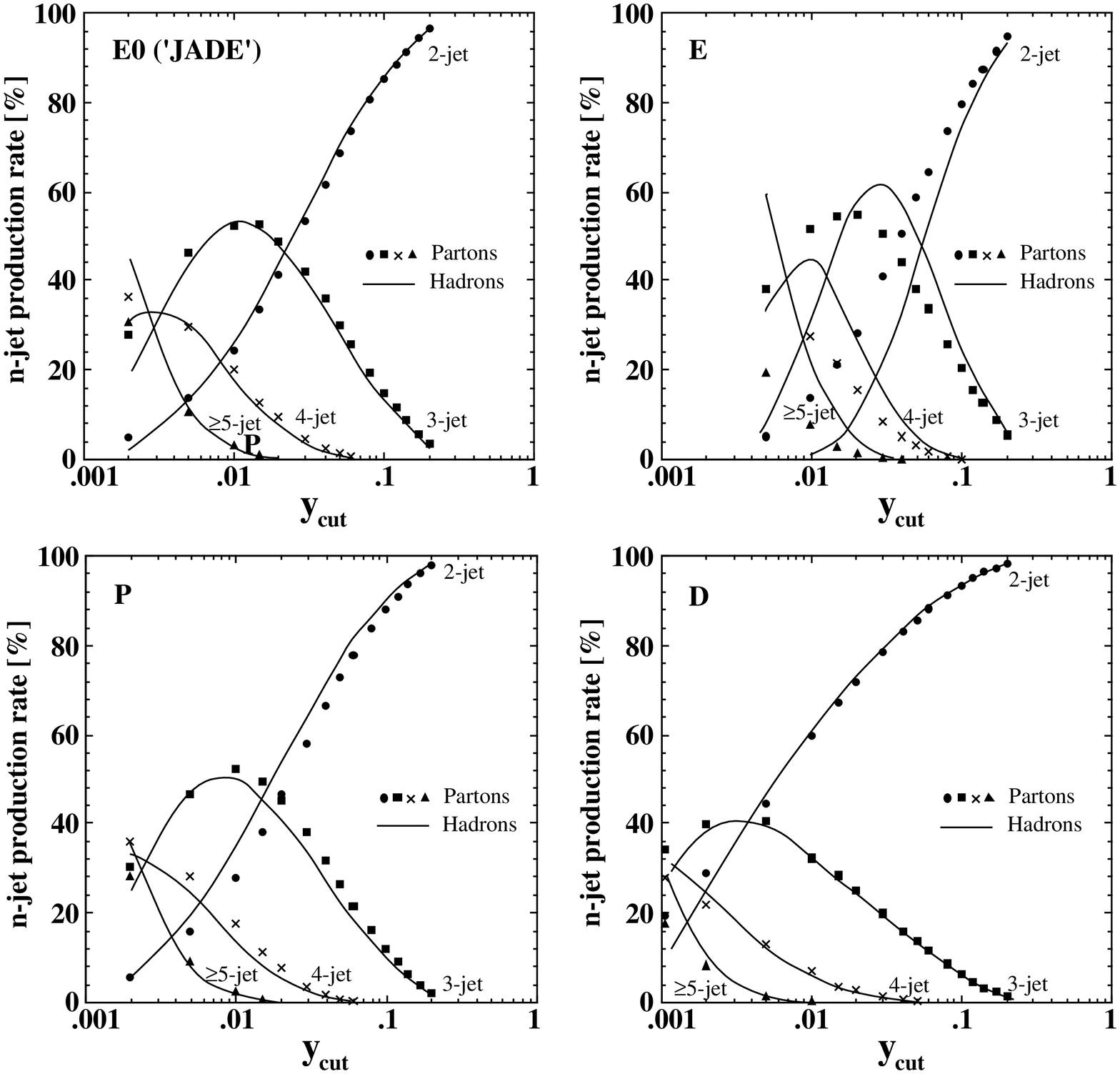,width=12.5cm}
\end{center}
\caption{\label{fig28}
}
\end{figure}

\begin{figure}[h,t]
\begin{center}
\epsfig{file=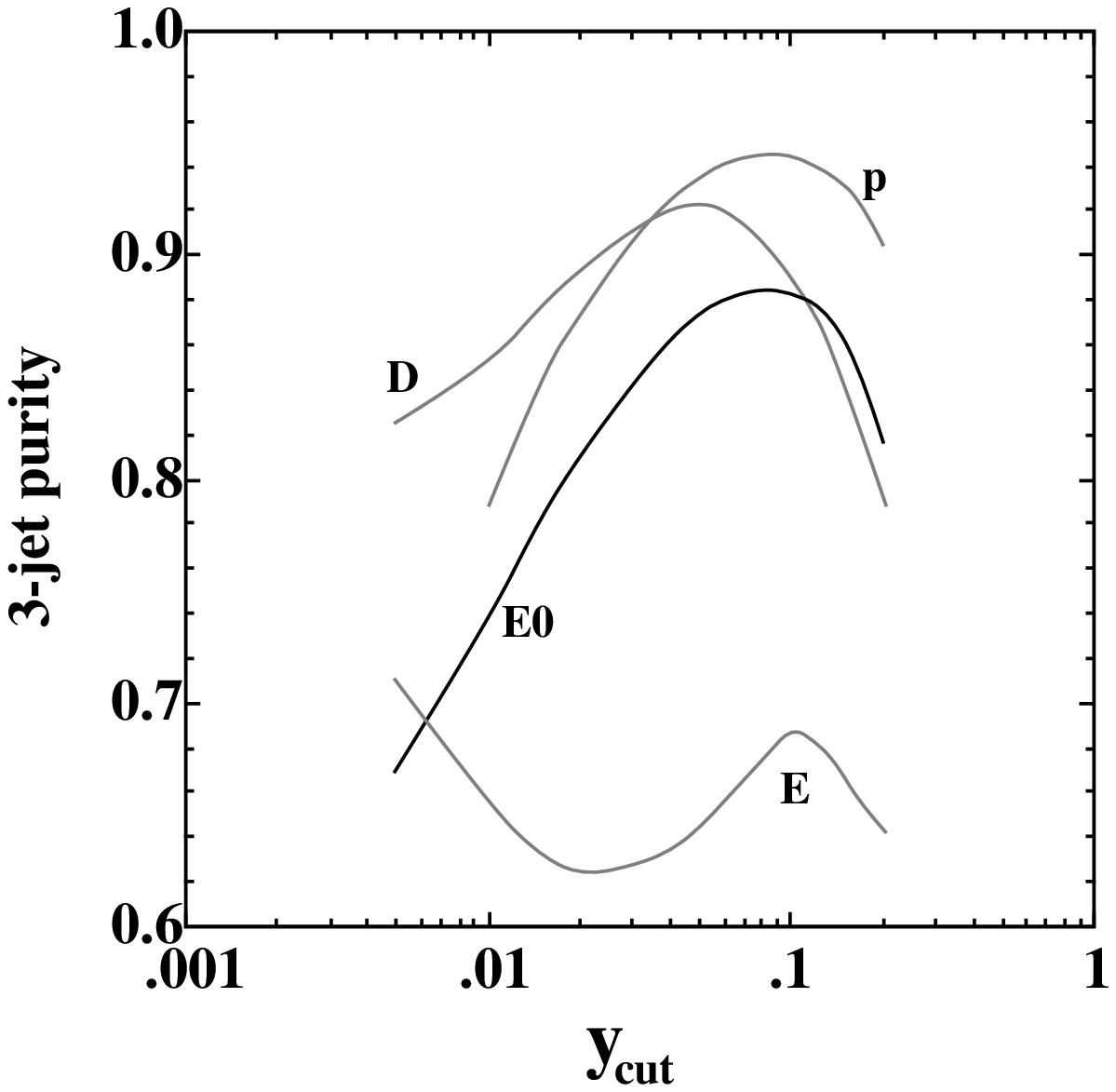,width=8.5cm}
\end{center}
\caption{\label{fig29}
}
\end{figure}

\begin{figure}[h,t]
\begin{center}
\epsfig{file=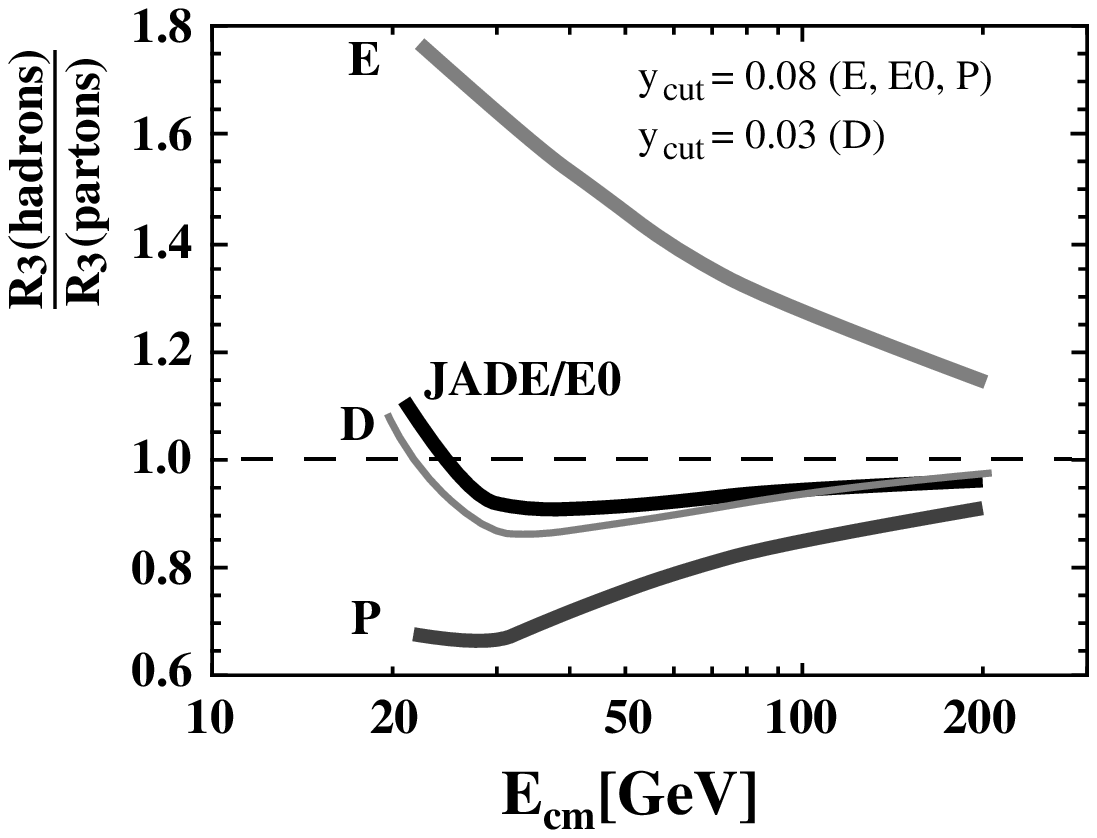,width=8.5cm}
\end{center}
\caption{\label{fig30}
}
\end{figure}

\begin{figure}[h,t]
\begin{center}
\epsfig{file=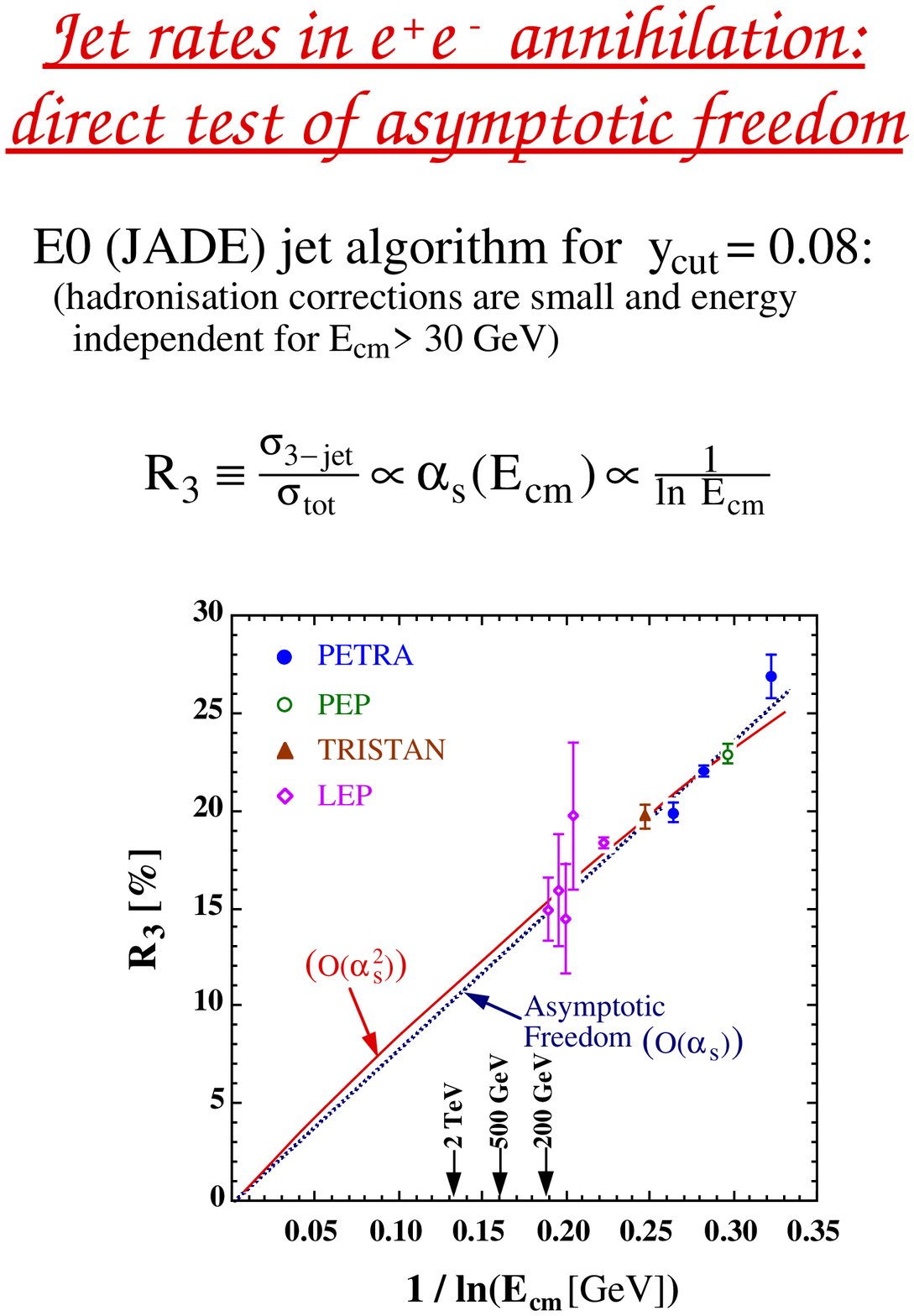,width=12.5cm}
\end{center}
\caption{\label{fig31}
}
\end{figure}

\begin{figure}[h,t]
\begin{center}
\epsfig{file=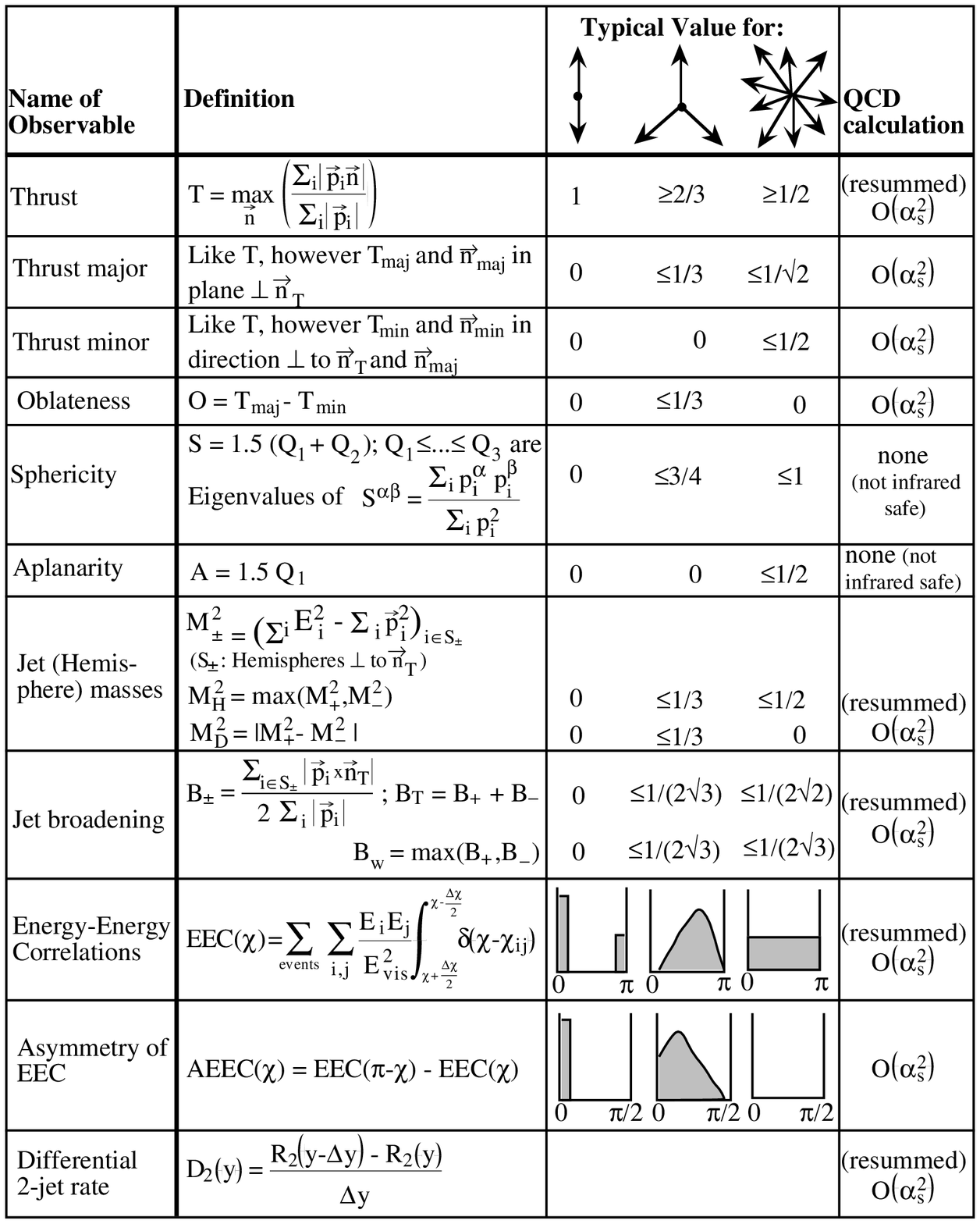,width=12.5cm}
\end{center}
\caption{\label{fig32}
}
\end{figure}

\begin{figure}[h,t]
\begin{center}
\epsfig{file=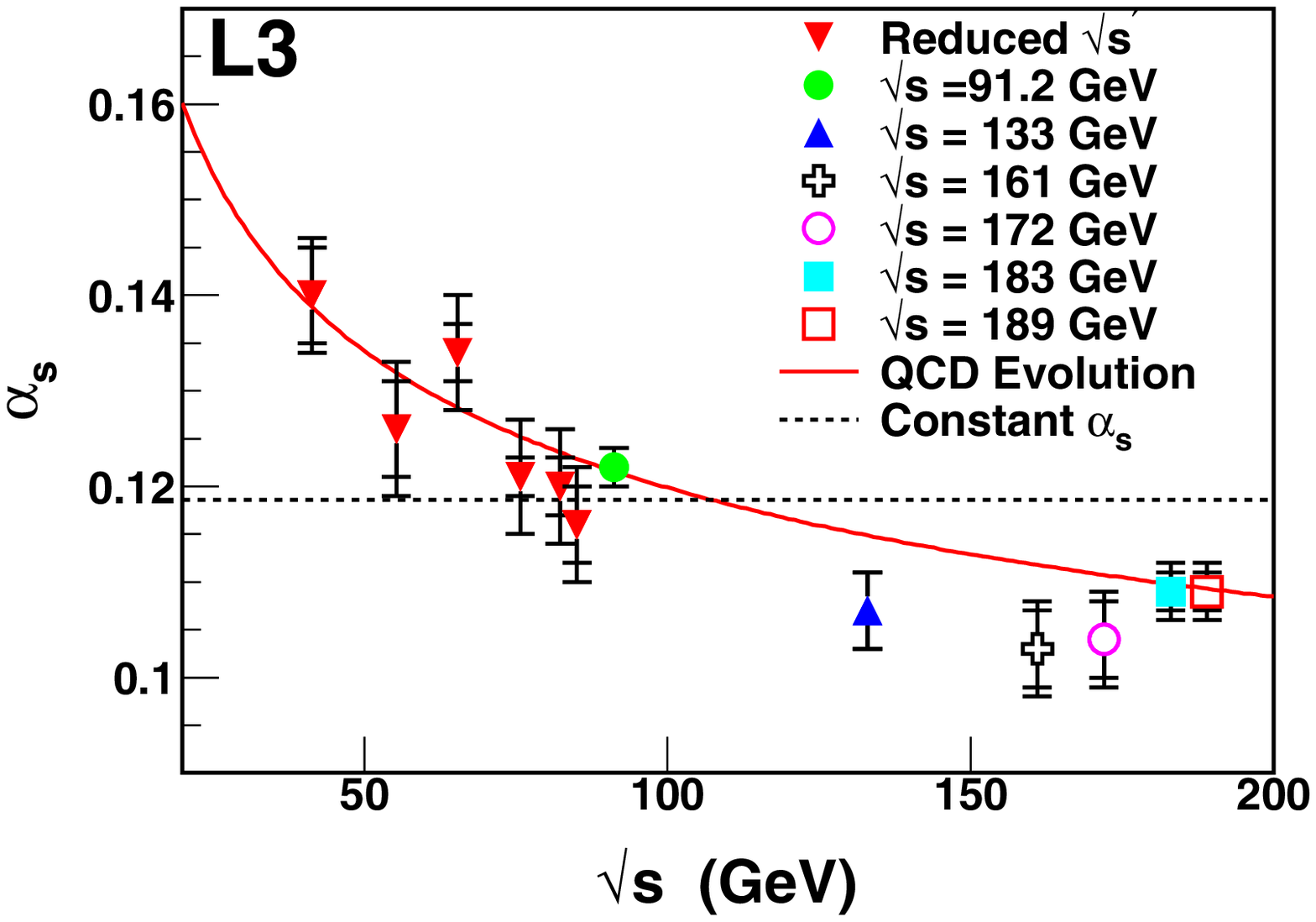,width=12.5cm}
\end{center}
\caption{\label{fig33}
}
\end{figure}

\begin{figure}[h,t]
\begin{center}
\epsfig{file=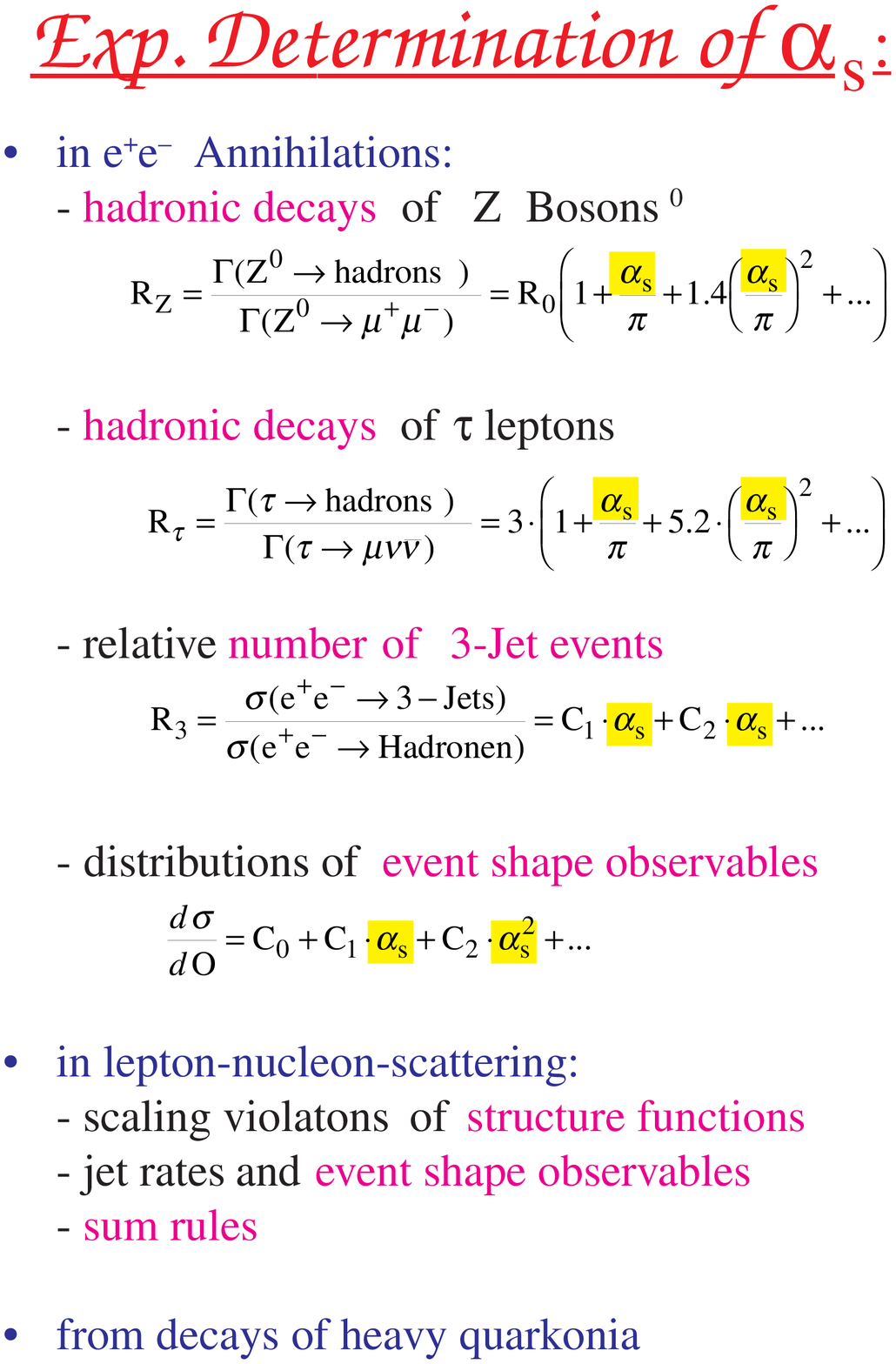,width=12.5cm}
\end{center}
\caption{\label{fig34}
}
\end{figure}

\begin{figure}[h,t]
\begin{center}
\epsfig{file=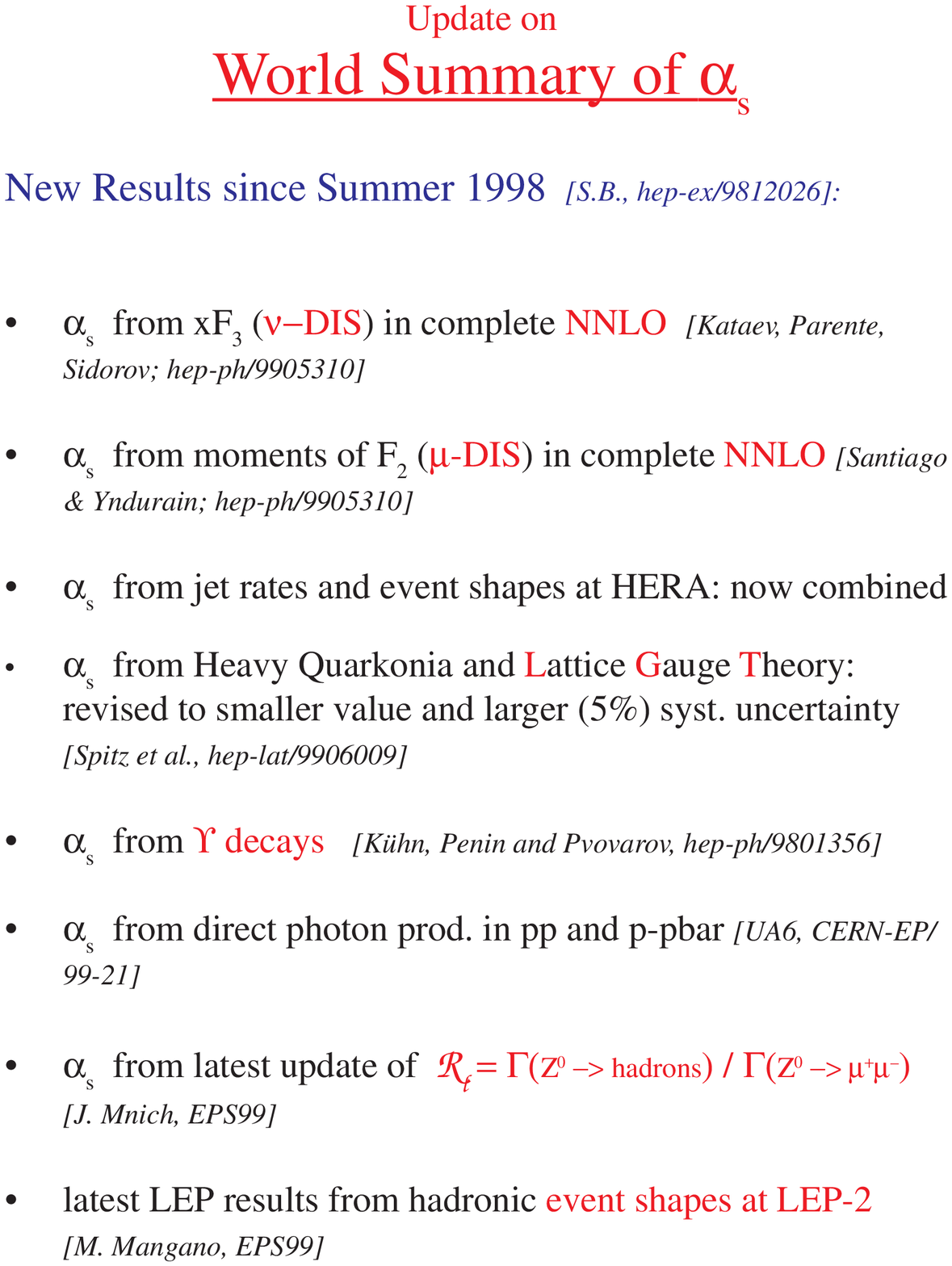,width=12.5cm}
\end{center}
\caption{\label{fig35}
}
\end{figure}
\clearpage

\renewcommand{\arraystretch}{1.3}
\begin{table}[h]
{\tiny
\begin{tabular}{|l|c|l|l|c c|c|}
   \hline 
  & Q & & &  \multicolumn{2}{c|}
{$\Delta \amz $} &  \\ 
Process & [GeV] & $\alpha_s(Q)$ &
  $ \amz$ & exp. & theor. & Theory \\
\hline \hline 
DIS [pol. strct. fctn.] & 0.7 - 8 & & $0.120\ ^{+\ 0.010}
  _{-\ 0.008}$ & $^{+0.004}_{-0.005}$ & $^{+0.009}_{-0.006}$ & NLO \\
DIS [Bj-SR] & 1.58
  & $0.375\ ^{+\ 0.062}_{-\ 0.081}$ & $0.121\ ^{+\ 0.005}_{-\ 0.009}$ & 
  -- & -- & NNLO \\
DIS [GLS-SR] & 1.73
  & $0.295\ ^{+\ 0.092}_{-\ 0.073}$ & $0.114\ ^{+\ 0.009}_{-\ 0.012}$ & 
  $^{+0.005}_{-0.006}$ & $^{+0.009}_{-0.010}$ & NNLO \\
$\tau$-decays 
  & 1.78 & $0.339 \pm 0.021$ & $0.121 \pm 0.003$
  & 0.001 &  0.003 & NNLO \\
\ul {DIS [$\nu$; ${\rm x F_3}$]}  & 5.0
  & $0.214 \pm 0.021$
   & $0.118\pm 0.006$   &
    $ 0.002 $ & $ 0.006$ & NNLO \\
\ul {DIS [e/$\mu$; ${\rm F_2}$]}
     & 2.96 & $0.252 \pm 0.011$ & $0.1172 \pm 0.0024$ & $ 0.0016$ &
     $ 0.0016$ & NNLO \\
\ul{DIS [e-p; jets \& shps]}
     & 7 - 100 &  & $0.118 \pm 0.006$ & $ 0.003$ &
     $0.005 $ & NLO \\
\ul {${\rm Q\overline{Q}}$ states}
     & 4.1 & $0.216 \pm 0.022$ & $0.115 \pm 0.006 $ & 0.000 & 0.006
     & LGT \\
\ul{$\Upsilon$ decays}
     & 4.75 & $0.217 \pm 0.021$ & $0.118 \pm 0.006
     $ & -- & -- & NLO \\
$\epem$ [$\sigma_{\rm had}$] 
     & 10.52 & $0.20\ \pm 0.06 $ & $0.130\ ^{+\ 0.021\ }_{-\ 0.029\ }$
     & $\ ^{+\ 0.021\ }_{-\ 0.029\ }$ & -- & NNLO \\
$\epem$ [jets \& shapes]  & 22.0 & $0.161\ ^{+\ 0.016}_{-\ 0.011}$ &
   $0.124\ ^{+\ 0.009}_{-\ 0.006}$ &  0.005 & $^{+0.008}_{-0.003}$
   & resum \\
$\epem$ [$\sigma_{\rm had}$]  & 34.0 &
 $0.146\ ^{+\ 0.031}_{-\ 0.026}$ &
   $0.123\ ^{+\ 0.021}_{-\ 0.019}$ & $^{+\ 0.021}_{-\ 0.019}
   $ & -- & NLO \\
$\epem$ [jets \& shapes] & 35.0 & $ 0.145\ ^{+\ 0.012}_{-\ 0.007}$ &
   $0.123\ ^{+\ 0.008}_{-\ 0.006}$ &  0.002 & $^{+0.008}_{-0.005}$
   & resum \\
$\epem$ [jets \& shapes] & 44.0 & $ 0.139\ ^{+\ 0.010}_{-\ 0.007}$ &
   $0.123\ ^{+\ 0.008}_{-\ 0.006}$ & 0.003 & $^{+0.007}_{-0.005}$
   & resum \\
$\epem$ [jets \& shapes]  & 58.0 & $0.132\pm 0.008$ &
   $0.123 \pm 0.007$ & 0.003 & 0.007 & resum \\
$\p\bar{\p} \rightarrow {\rm b\bar{b}X}$
    & 20.0 & $0.145\ ^{+\ 0.018\ }_{-\ 0.019\ }$ & $0.113 \pm 0.011$ 
    & $^{+\ 0.007}_{-\ 0.006}$ & $^{+\ 0.008}_{-\ 0.009}$ & NLO \\
\ul {${\rm p\bar{p},\ pp \rightarrow \gamma X}$}  & 24.2 & $0.138
 \ ^{+\ 0.011}_{-\ 0.009}$ &
  $0.111\ ^{+\ 0.008\ }_{-\ 0.005\ }$ & 0.002 &
  $^{+\ 0.008}_{-\ 0.005}$ & NLO \\
${\sigma (\rm p\bar{p} \rightarrow\  jets)}$  & 30 - 500 &  &
  $0.121\pm 0.009$ & 0.001 & 0.009 & NLO \\
\ul{$\epem$ [$\Gamma (\z0 \rightarrow {\rm had.})$]}
    & 91.2 & $0.123\pm 0.005$ & 
$0.123\pm 0.005$ &
   $ 0.004$ & $0.003$ & NNLO \\
$\epem$ [jets \& shapes] &
    91.2 & $0.122 \pm 0.006$ & $0.122 \pm 0.006$ & $ 0.001$ & $
0.006$ & resum \\
$\epem$ [jets \& shapes]  & 133.0 & $0.111\pm 0.008$ &
   $0.117 \pm 0.008$ & 0.004 & 0.007 & resum \\
$\epem$ [jets \& shapes]  & 161.0 & $0.105\pm 0.007$ &
   $0.114 \pm 0.008$ & 0.004 & 0.007 & resum \\
$\epem$ [jets \& shapes]  & 172.0 & $0.102\pm 0.007$ &
   $0.111 \pm 0.008$ & 0.004 & 0.007 & resum \\
$\epem$ [jets \& shapes]  & 183.0 & $0.109\pm 0.005$ &
   $0.121 \pm 0.006$ & 0.002 & 0.006 & resum \\
\ul{$\epem$ [jets \& shapes]} & 189.0 & $0.110\pm 0.004$ &
   $0.123 \pm 0.005$ & 0.002 & 0.005 & resum \\
& & & & & & \\
\hline
\end{tabular}
\caption{
World summary of measurements of $\as$.
Underlined entries 
are new or updated since autumn 1998
(DIS = deep inelastic scattering; GLS-SR = Gross-Llewellyn-Smith sum rules;
Bj-SR = BjorkLG sum rttes;
(N)NLO = (next-to-)next-to-leading order perturbation theory;
LGT = lattice gauge theory;
resum. = resummed next-to-leading order). \label{astab}}
}
\end{table}
\clearpage

\begin{figure}[h,t]
\begin{center}
\epsfig{file=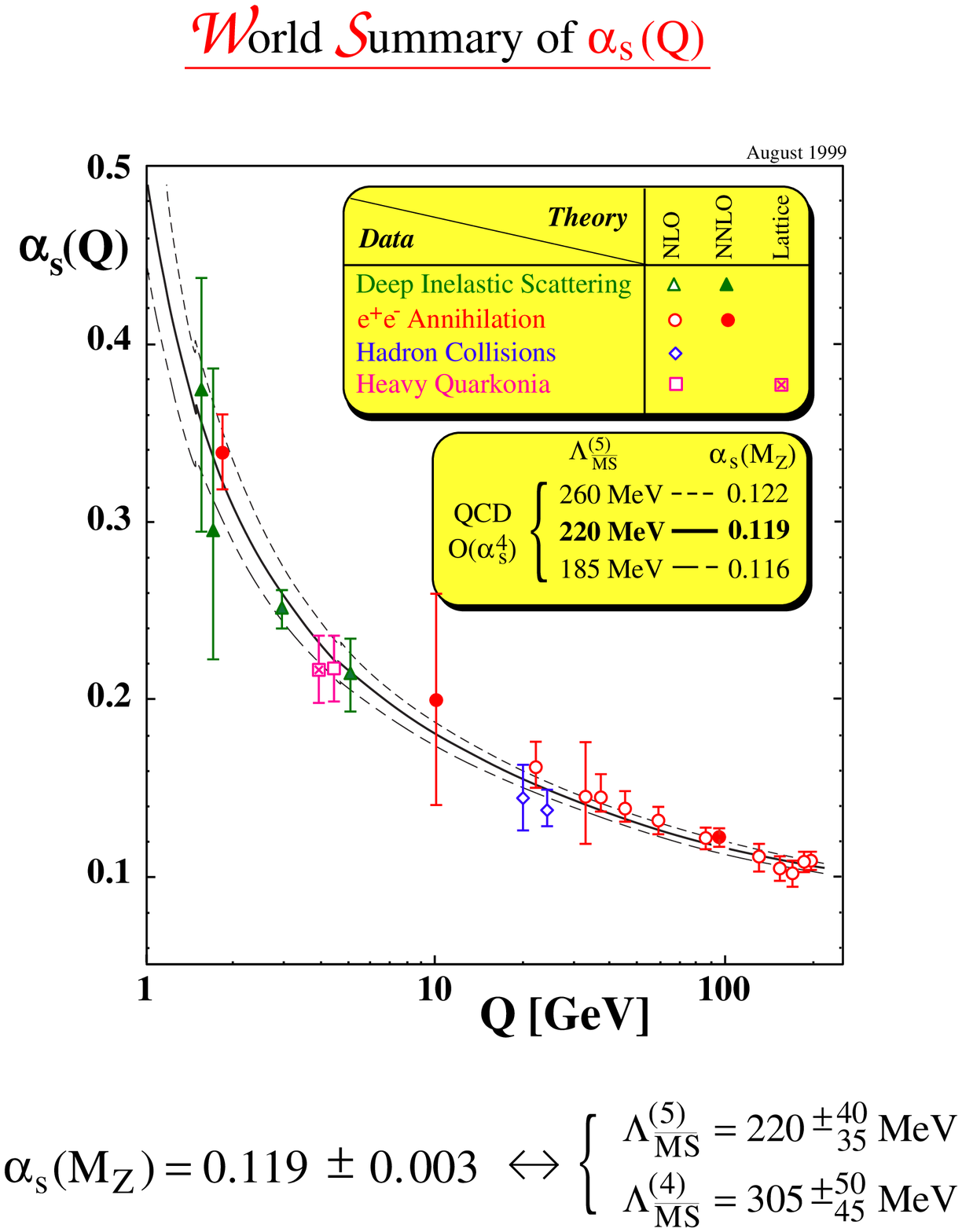,width=12.5cm}
\end{center}
\caption{\label{fig37}
}
\end{figure}

\begin{figure}[h,t]
\begin{center}
\epsfig{file=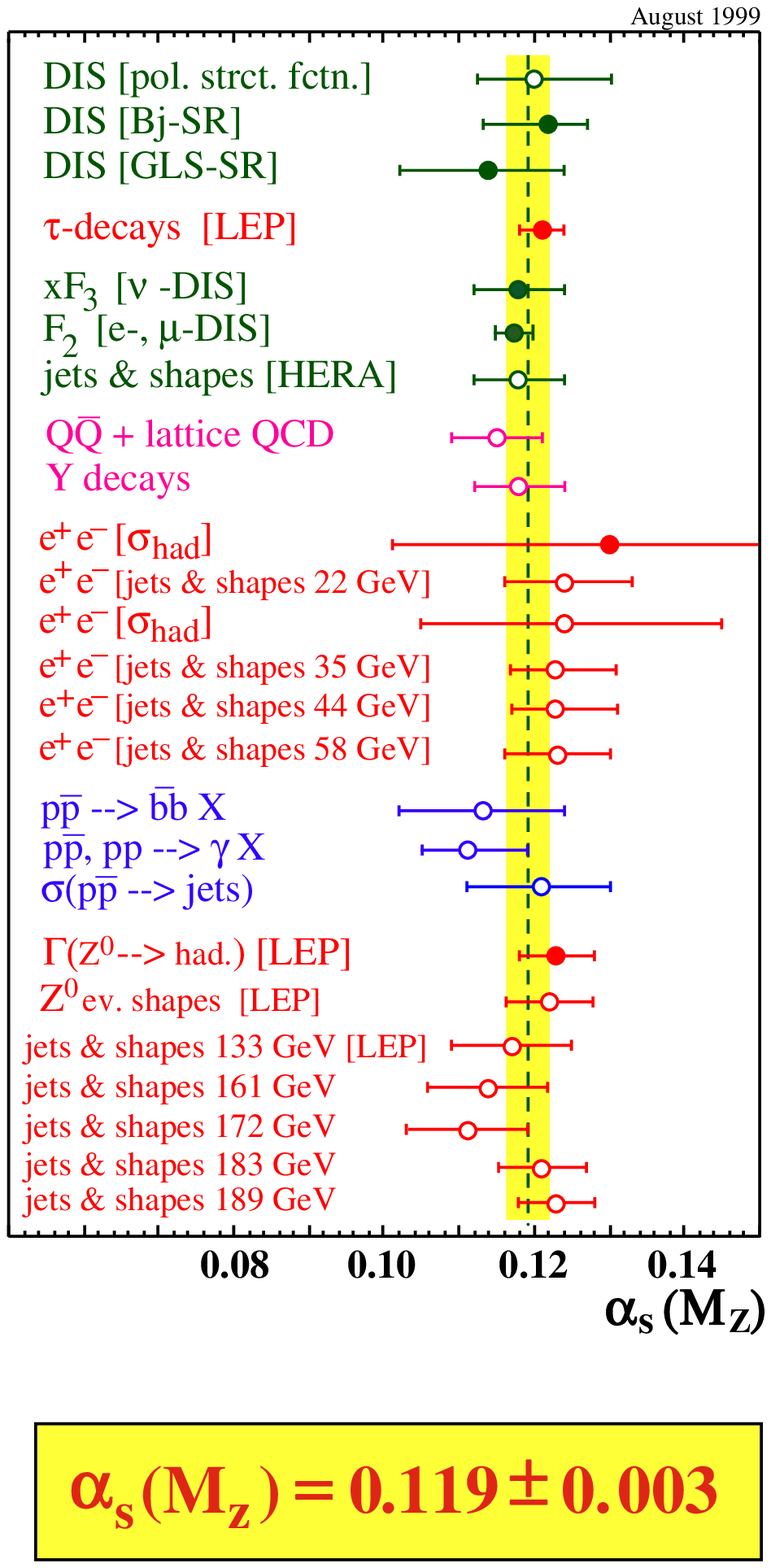,height=18.3cm}
\end{center}
\caption{\label{fig38}
}
\end{figure}

\end{document}